\documentclass[aps,prd,a4paper,longbibliography,reprint,superscriptaddress,floatfix,nofootinbib,amsmath,amssymb]{revtex4-1}
\usepackage{color,hhline,psfrag,rotating}
\usepackage{dcolumn}
\usepackage{bm}
\usepackage[utf8]{inputenc}
\usepackage{hyperref}
\newcommand{\I}{\mathrm{i}}
\begin{document}
\title{$\rho$ and $K^*$ resonances on the lattice 
at nearly physical quark masses and $N_f=2$}

\author{Gunnar~S.~Bali}
\affiliation{Institute for Theoretical Physics, University of Regensburg, 93040 Regensburg, Germany}
\affiliation{Tata Institute of Fundamental Research, Homi Bhabha Road, Mumbai 400005, India}
\author{Sara~Collins}
\affiliation{Institute for Theoretical Physics, University of Regensburg, 93040 Regensburg, Germany}
\author{Antonio~Cox}
\affiliation{Institute for Theoretical Physics, University of Regensburg, 93040 Regensburg, Germany}
\author{Gordon~Donald}
\affiliation{Institute for Theoretical Physics, University of Regensburg, 93040 Regensburg, Germany}
\author{Meinulf~Göckeler}
\affiliation{Institute for Theoretical Physics, University of Regensburg, 93040 Regensburg, Germany}
\author{C.~B.~Lang}
\affiliation{Institute of Physics, University of Graz, A-8010 Graz, Austria}
\author{Andreas~Schäfer}
\affiliation{Institute for Theoretical Physics, University of Regensburg, 93040 Regensburg, Germany}
\collaboration{RQCD Collaboration}
\date{\today}
\begin{abstract}
Working with a pion mass $m_\pi \approx 150\,\textmd{MeV}$, we study $\pi\pi$ and $K\pi$
scattering using two flavours of non-perturbatively improved Wilson fermions
at a lattice spacing $a\approx 0.071\,\textmd{fm}$. 
Employing two lattice volumes with linear spatial extents of
$N_s=48$ and $N_s=64$ points
and moving frames, we extract the phase shifts for $p$-wave
$\pi\pi$ and $K\pi$ scattering near the $\rho$ and $K^*$ resonances.
Comparing our results
to those of previous lattice studies, that used pion masses ranging
from about $200\,\textmd{MeV}$ up to $470\,\textmd{MeV}$,
we find that the coupling $g_{\rho\pi\pi}$
appears to be remarkably constant as a function of $m_{\pi}$.
\end{abstract}
\maketitle
\section{Introduction}
\label{sec:intro}
Lattice QCD calculations are particularly suited for studies of hadrons
which are stable under the strong interaction and their properties
can be determined by studying correlation functions
at large Euclidean time separations.
However, almost all known hadrons are unstable resonances, which
complicates the situation. The $\rho$ meson, one of the simplest
resonances in QCD, couples to a pair of pions with total isospin $I=1$.
In a finite lattice volume of linear spatial size $L=N_sa$, the allowed momenta of the pion pair are quantized. Neglecting $\pi\pi$ interactions,
the lowest lying $\pi\pi$ state with
the same spin $J=1$ as the $\rho$ has the energy
\begin{equation}
\label{eq:energy}
E_{\pi\pi}^{\mathrm{free}} = 2 \sqrt{ m_\pi^2 + \left(\frac{2\pi}{L}\right)^2 }\,.
\end{equation}
The $\rho$ can only be treated as a stable particle if its mass
is sufficiently smaller than this $\pi\pi$ centre of momentum frame 
energy $E_{\pi\pi}^{\mathrm{free}}$. This is possible if
the pion is heavy or the lattice size is small.
For the values of $m_\pi$ and $L$ that are now accessible in lattice
simulations this is not the case anymore.

The formalism for dealing with resonances in lattice QCD simulations of two-particle scattering systems has been developed first with equal masses and in systems at rest \cite{Luscher:1990ux,Luscher:1986pf}
and later extended to various moving frames and unequal masses \cite{Rummukainen:1995vs,Kim:2005gf,Christ:2005gi,Feng:2010es,Davoudi:2011md,Fu:2011xz,Leskovec:2012gb,Gockeler:2012yj}. 
In $\pi\pi$ scattering, the $\rho$ appears as an increase of the scattering phase shift from zero to $\pi$ as the centre of momentum frame energy, $E_{cm}$,
is varied from below to above the resonant mass value $m_{\rho}$.
The dependence of the $\ell=1$ angular momentum partial wave shift
$\delta$ on $E_{cm}$ gives detailed information about the nature of the resonance. To first approximation, the resonant mass can be extracted at the
value $\delta=\pi/2$.

Due to the computational cost, previous calculations of the resonance parameters
were restricted to unphysically large pion masses (most even employed pion masses
with $E_{\pi\pi}^{\mathrm{free}}>m_{\rho}$),
but the expected phase shift behaviour was still observed \cite{Aoki:2007rd,Feng:2010es,Frison:2010ws,Lang:2011mn,Aoki:2011yj,Pelissier:2012pi,Dudek:2012xn,Fahy:2014jxa,Wilson:2015dqa,Bulava:2015qjz,Guo:2015dde}. Algorithmic advances and
increases in compute power now enable us
to pursue the first scattering study at a close to physical pion mass
$m_{\pi}\approx 150\,\textmd{MeV}$.

The strange-light analogue of the light-light
$\rho$ meson is the $K^*$. Its phase shift has also been studied previously
in lattice calculations at unphysically large pion masses \cite{Fu:2012tj,Prelovsek:2013ela,Dudek:2014qha,Wilson:2014cna}.
There are similarities between $\pi\pi$ and $K\pi$ scattering
not only in terms of the formalism but also
in terms of constructing and
computing the necessary correlation functions, which means we
can incorporate the $K^*$ resonance into our study,
with limited computational overhead.

\begin{table*}
\caption{Details of the lattice configurations: volume, coupling,
lattice spacing (determined in Ref.~\cite{Bali:2012qs}), light and
strange quark mass
parameters $\kappa_{\ell}$ and $\kappa_s$, (finite volume) pion mass, kaon mass, the linear spatial
size in units of the infinite volume
pion mass $Lm_{\pi}^{\infty}$~\cite{Bali:2014gha}, the unit momentum
$2\pi / L$ and the number of configurations $N_{\mathrm{cfg}}$
analysed. The errors given for $m_{\pi}$ and $m_K$ are statistical
only and do not include the 3\% scale setting uncertainty~\protect\cite{Bali:2012qs}.}
\label{tab:lattices}
\centering
\begin{ruledtabular}
\begin{tabular}{cccccccccc} 
$N_s^3 \times N_t$ & $\beta$ & $a^{-1}$ & $\kappa_{\ell}$ & $\kappa_s$ & $m_\pi$ & $m_K$ & $Lm_\pi^{\infty}$ & $2\pi/L$ & $N_{\mathrm{cfg}}$ \\
\hline
$48^3 \times 64$ & 5.29 & 2.76(8) \textmd{GeV} & 0.13640 & 0.135574 & 160(2) \textmd{MeV} & 500(1) \textmd{MeV} & 2.61 & 361 \textmd{MeV} & 888 \\
$64^3 \times 64$ & 5.29 & 2.76(8) \textmd{GeV} & 0.13640 & 0.135574 & 150(1) \textmd{MeV} & 497(1) \textmd{MeV} & 3.48 & 271 \textmd{MeV} & 671 \\
\end{tabular}
\end{ruledtabular}
\end{table*}

From experiment, the $\rho$ has a mass of around $775\,\textmd{MeV}$
and a decay width
$\Gamma_{\rho}\approx 148\,\textmd{MeV}$ while the $K^*$ mass and width
are approximately $896\,\textmd{MeV}$ and $47\,\textmd{MeV}$~\cite{pdg},
respectively.
The decays are almost exclusively to $\pi\pi$ and $K\pi$.
In our study of the $\rho$ resonance we neglect couplings to three- and
four-pion states. 
Our calculation (and all other $\pi\pi$ scattering calculations to-date)
is performed with isospin symmetry in place, therefore
$3\pi$ final states are excluded.
Isospin symmetry tremendously simplifies the computation
for the $I=1$ $\rho$ and the $I=1/2$ $K^*$ channels we consider here
as there are no disconnected quark-line contractions.
As we will see, at our pion mass and for the kinematics
we implement, only one of our data points could be sensitive to
$4\pi$ final states. Also, considering the available phase space and
Okubo-Zweig-Iizuka suppression, neglecting
these multi-particle final states should be a very good approximation.
This argument is supported by experimental evidence,
indeed suggesting a virtually undetectable
coupling of the $\rho$ meson to $4\pi$ states \cite{Akhmetshin:1999ty}.
Comparing measurements of the branching fractions of $\rho\to 4\pi$ and the (isospin breaking) $\rho\to 3\pi$ decay \cite{Akhmetshin:1999ty,Achasov:2003ir} shows that they are of similar (small) sizes.
For a neutral $\rho$ meson, the decay width to $\pi^+\pi^-\pi^0$ is $15(7)\,\textmd{keV}$.
Combining the widths to $\pi^+\pi^-\pi^0\pi^0$ and $\pi^+\pi^-\pi^+\pi^-$ gives
$5(2)\,\textmd{keV}$. This is indeed negligible, relative to the
total width of $148\,\textmd{MeV}$. For decays of a charged
$\rho$ into four pions
only an upper limit exists. 

In the cases of $\pi\pi$ and $K\pi$ scattering, respectively,
in principle there could also
be interference with $K\overline K$
and $K\eta$; $2m_K\approx 985\,\textmd{MeV}$,
$m_K + m_\eta \approx 1040\,\textmd{MeV}$. However, both values are
well above the region we are interested in, in particular
considering $p$-wave decay in a finite volume.
For heavier than physical pions, these thresholds
are closer. This situation was studied
at $m_\pi \approx 236\,\textmd{MeV}$ in Ref.~\cite{Wilson:2015dqa}
for the $\rho$ resonance
and at $m_\pi \approx 391\,\textmd{MeV}$
in Ref.~\cite{Wilson:2014cna} for the $K^*$.
Indeed, even at these large pion masses,
the impact was found to be negligible.
Finally, we also ignore $K^*\rightarrow K\pi\pi$, noting that the
upper limit reads
$\Gamma(K^* \to K\pi\pi)\approx 35\,\textmd{keV}$ \cite{Jongejans:1977ty};
the vast majority of experimentally observed decays to $K\pi\pi$ final states
appear to be related to heavier resonances \cite{Aston:1986jb}.

Our method to generate the necessary correlation functions
has been employed in previous
calculations~\cite{Aoki:2007rd,Feng:2010es,Aoki:2011yj}. 
Nevertheless, we provide a brief description
of the construction of correlators, along with details on the
lattices and kinematics used in Sec.~\ref{sec:calc}.
The results are presented and discussed in Sec.~\ref{sec:results},
before we conclude in Sec.~\ref{sec:conclude}.

\section{Lattice calculation}
\label{sec:calc}
We aim to extract the resonance parameters (mass and width) of the $\rho$ and $K^*$ from their appearances in $\pi\pi$ and $K\pi$ $p$-wave scattering, respectively.
To do so, we will determine the spectra of interacting two-particle QCD states in finite volumes.
Using these energy levels, along with known relations, allows us to extract the scattering phase shift, from whose dependence on the
energy $E_{cm}$ in the rest frame of the $\rho$ (or the $K^*$)
the resonance parameters can be found.

\subsection{Discussion of the lattice parameters}
We employ lattice configurations with a lattice spacing
$a\approx 0.071\,\textmd{fm}$ and time extent $N_ta=64a\approx4.6\,\textmd{fm}$,
generated by the Regensburg lattice QCD group (RQCD, $L=64a$) and
RQCD/QCDSF ($L=48a$)
with $N_f = 2$ flavours of degenerate non-perturbatively improved Wilson
sea quarks with a pion mass of about $150\,\textmd{MeV}$
(ensembles VIII and VII of
Ref.~\cite{Bali:2014nma}). On the larger volume every second
trajectory and on the smaller volume every fifth trajectory is analysed.
Discretization errors are of $\mathcal{O}(a^2)$. We expect these to be small for
the light hadron masses considered at our
lattice scale $a^{-1}= 2.76(8)\,\textmd{GeV}$~\cite{Bali:2012qs}.
The lattice parameters are given in Table~\ref{tab:lattices}.
More detail can be found in Refs.~\cite{Bali:2014nma,Bali:2014gha}.
Following Ref.~\cite{Bali:2011ks},
we check the strange quark mass tuning by computing
$\sqrt{2m_K^2 - m_\pi^2}=686.5(1.1)\,\textmd{MeV}$
on the $N_s=64$ ensemble, assuming $a^{-1}=2.76\,\textmd{GeV}$.
We find perfect agreement with
the ``experimental'' value of $686.9\,\textmd{MeV}$.

The choice of our ensembles is motivated by the
proximity of the pion mass to its experimental value.
In the $\rho\rightarrow\pi\pi$ channel the pions must have
relative angular momentum.
For a system at rest this is only possible
if their individual momenta are non-zero. This gives
the threshold Eq.~\eqref{eq:energy}, where
$m_{\rho}>E_{cm}^{\mathrm{free}}>2m_{\pi}$,
for the $\rho$ to become unstable in a finite volume.
On our lattice configurations, this threshold
lies at $782\,\textmd{MeV}$ (within the experimental $\rho$ resonance width) for
$N_s=48$ and at $619\,\textmd{MeV}$ (beneath the resonance) for $N_s=64$.
Note that in moving frames the effective thresholds can be lower. 

The combination $Lm_{\pi}$ is the relevant quantity controlling
finite size effects. This combination
obviously decreases with $m_\pi$ and it is expensive to enlarge the linear
box size $L$ to fully compensate for this.
Our lattice volumes have $Lm_{\pi}<4$, due to limited
computer resources. However, there are clear
advantages to employ small volumes for resolving broad resonances
like the $\rho$:
At large $L$ the spectrum of two-particle
states becomes dense, complicating the extraction of the relevant
energy levels and increasing the demand on the precision of their
determination. 

Terms which are exponentially
suppressed in $Lm_\pi$ are neglected
in the L{\"u}scher phase shift method~\cite{Luscher:1990ux}.
One such effect is the difference between
the pion mass $m_{\pi}\approx 160\,\textmd{MeV}$ on the small volume
and its infinite volume value
$m_{\pi}^{\infty}\approx 149.5\,\textmd{MeV}$~\cite{Bali:2014nma},
which goes beyond this formalism.
Note that
$Lm_{\pi}^{\infty}\approx 2.6$ for our smaller volume and $e^{-2.6}\approx 0.074$
may not necessarily be considered a small number. 
Fortunately, it has been demonstrated, at least in some models, e.g.,
in the inverse amplitude
and the $N/D_A$ models, that for
$I=1$ $p$-wave $\pi\pi$ scattering the corrections to the L\"uscher
formula may be negligible as long as $Lm_{\pi}>2$~\cite{Albaladejo:2013bra}.
We note that towards small pion masses the $\rho$ resonance broadens,
allowing us to extract non-trivial phase shifts for
a wider range of energies than had
been possible in previous simulations at unphysically large
pion masses. This allows us to collect several data points
within the region relevant to constrain the resonance parameters.

An issue that arises for pions which are sufficiently close to their
physical mass is the opening of the four-pion threshold as, in
nature, $m_\rho > 4m_\pi$.
In analogy to our discussion of two-particle thresholds, we can determine
where the four-particle thresholds will lie for the lattice configurations
we use.
When the $\rho$ meson is at rest at least two of the pions need
to carry non-zero momenta. 
In this case, a decay to four pions requires $918\,\textmd{MeV}$ on our larger
lattice size and $1081\,\textmd{MeV}$ on the smaller one, both of which
lie well above the resonance region.

Again, for moving frames, these limits can be lower.
We encounter the worst case for the total momentum
$\mathbf{P} = (0,0,1) (2\pi/L)$ on $L=64a$, where the four-pion threshold lies
around $E_{cm} = 710\,\textmd{MeV}$.
Fortunately, as we discussed in the introduction,
the $\rho$ and $K^*$ resonances are entirely dominated by $p$-wave decays
into $\pi\pi$ and $K\pi$ final states; even in experiment
other channels are hardly detectable at all.
Finally, we remark that dealing with decays to more than two particles
in lattice QCD is an open problem. While there has been recent
theoretical progress addressing
three-particle final states
\cite{Kreuzer:2009jp,Briceno:2012rv,Hansen:2014eka,Meissner:2014dea,Hansen:2015zga}, we do not know how to analyse four-pion states in a lattice calculation.

\subsection{Generation of the correlators}
In order to treat the $\rho$ as a resonance in $\pi\pi$ scattering, we employ
a basis of interpolators which explicitly couple to one- and two-particle states.
The interpolators used for each kinematic setting
all share the same quantum numbers and symmetries.
In the case of $\pi\pi$ scattering, we are interested in
the $I=1$, $J^{P}=1^-$ channel in which the $\rho$ appears.
The $\pi\pi$ interpolators read
\begin{equation}
\pi(\mathbf{p}_1)\pi(\mathbf{p}_2)=\frac{1}{\sqrt{2}}\left[
\pi^+(\mathbf{p}_1)\pi^-(\mathbf{p}_2) - \pi^-(\mathbf{p}_1)\pi^+(\mathbf{p}_2)\right]\,,
\end{equation}
where $\pi=\bar{\psi}\gamma_5\psi$ and
the one-particle vector interpolator has the momentum $\mathbf{P}=\mathbf{p}_1+\mathbf{p}_2$.
For this we use
three structures in our basis:
$\bar{\psi} \gamma_j \psi$, $\bar{\psi} \gamma_j \gamma_t \psi$ and
$\bar{\psi} \nabla_j \psi$.

We apply Wuppertal quark smearing \cite{Gusken:1989ad}, where the field, $\phi^{(n)}_x$, at site $x$ after $n$ smearing iterations is 
\begin{equation}
\label{eq:wuppertal}
\phi^{(n)}_x=\frac{1}{1+6\delta}\left(\phi^{(n-1)}_x+\delta\sum_{j=\pm 1}^{\pm 3}U_{x,j}\phi^{(n-1)}_{x+a\hat{\boldsymbol{\jmath}}}\right).
\end{equation}
We set $\delta = 0.25$ and employ three levels of quark smearing,
using 50, 100 or 150 iterations. $U_{x,\mu}$ is a (smeared) gauge link
connecting $x$ with $x+a\hat{\mu}$ and
$U_{x,-\mu}=U_{x-a\hat{\mu},\mu}^{\dagger}$.
For the pseudoscalar meson operators, we choose the narrowest smearing width.
We use all three smearing levels for $\bar{\psi} \gamma_j \psi$ and $\bar{\psi} \gamma_j \gamma_t \psi$ and only the narrowest for $\bar{\psi} \nabla_j \psi$, so we have one two-particle interpolator and a total of seven one-particle
interpolators.
We employ spatial APE smearing for the gauge links \cite{Falcioni:1984ei} that
appear within Eq.~\eqref{eq:wuppertal} above:
\begin{equation}
\label{eq:smear}
U_{x,i}^{(n)}= P_{\mathrm{SU}(3)}\!\left(\!\alpha\,U_{x,i}^{(n-1)}+\sum_{|j|\neq i}
U_{x,j}^{(n-1)}U^{(n-1)}_{x+a\hat{\boldsymbol{\jmath}},i}U^{(n-1)\dagger}_{x+a\hat{\boldsymbol{\imath}},j}\!\right)
\end{equation}
with $i\in\{1,2,3\}, j\in\{\pm 1,\pm 2,\pm 3\}$. 
$P_{\mathrm{SU}(3)}$ denotes a projection into the $\mathrm{SU}(3)$ group. 
We use $\alpha = 2.5$ and 25 iterations.

\begin{figure}
\centering
\includegraphics[width=0.48\textwidth]{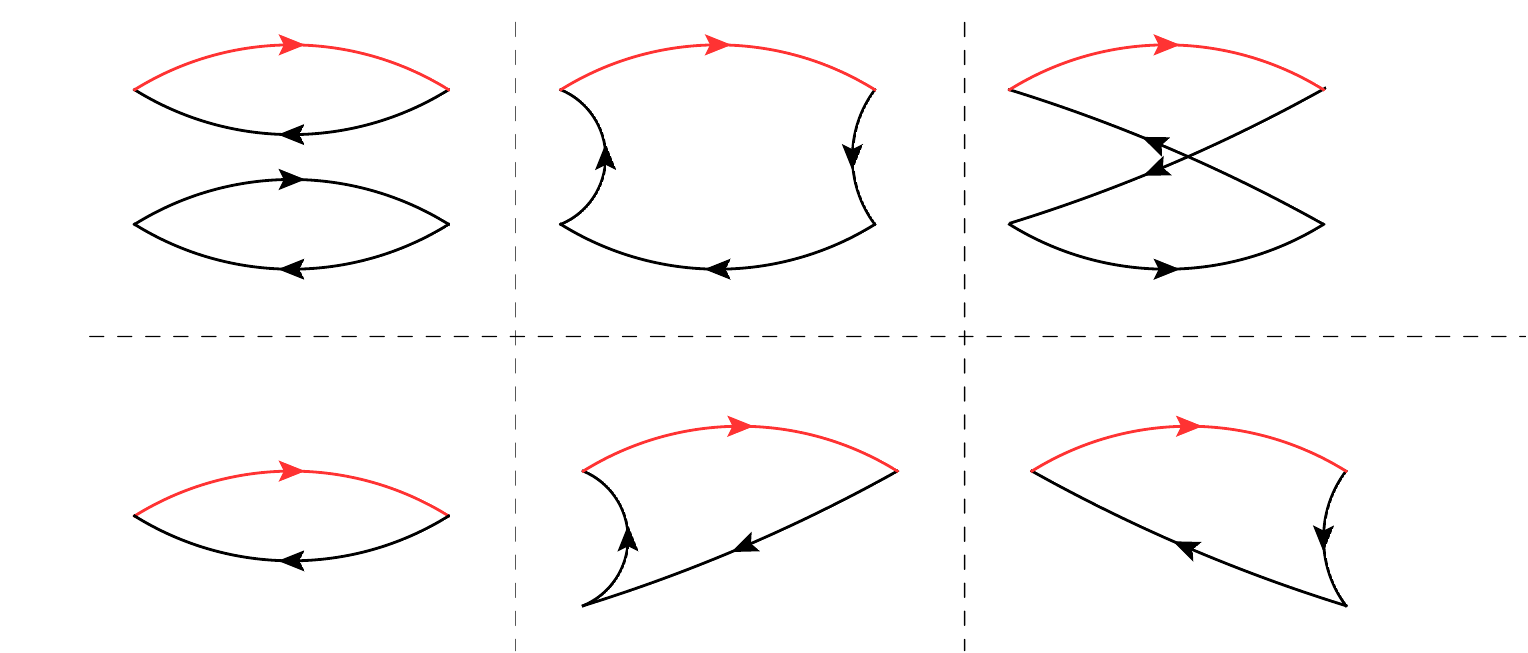}
\caption{Diagrams showing the quark contractions for the entries of our correlation matrices for $\pi\pi$ and $K\pi$ scattering. The red propagator denotes a strange quark for $K\pi$ and a light one for $\pi\pi$, while the black propagators are for light quarks in both cases. The top row shows contractions for $2\to2$ particle correlators. The right-most diagram does not appear for $I=1$ $\pi\pi$ scattering, but is required for our $K\pi$ scattering calculation. The second row contains $1\to 1$, $2\to 1$ and $1\to 2$ entries, respectively, where the $1\to 2$ element is the complex conjugate of $2\to 1$.}
\label{fig:contract}
\end{figure}

In $K\pi$ scattering, the $K^*$ resonance is in the $I=1/2$ channel, so we
use
\begin{equation}
\pi(\mathbf{p}_1)K(\mathbf{p}_2)=\sqrt{\frac{2}{3}} \pi^+(\mathbf{p}_1) K^-(\mathbf{p}_2) - \sqrt{\frac{1}{3}} \pi^0(\mathbf{p}_1) K^0(\mathbf{p}_2)
\end{equation}
as the two-particle interpolator.
The one-particle interpolators are the same as for the $\rho$ resonance,
replacing one light quark by the strange.
From these interpolators we calculate a matrix of correlation functions.
The contractions for its entries are depicted in Fig.~\ref{fig:contract}.

\begin{table*}
\caption{The interpolators we use in different moving frames. For each frame, we note the little group and irreducible representations (irrep) we employ and the one- and two-particle interpolators that belong to them. $\mathbf{K}$
denotes the integer valued total momentum
vector (used for the one-particle interpolator) and the arguments of
the pseudoscalar interpolators are $\mathbf{k}_j=\mathbf{p}_jL/(2\pi)$.}
\label{tab:irreps}
\centering
\begin{ruledtabular}
\begin{tabular}{cccccc}
$\pi\pi$&&&& \\
$N_s$&$\mathbf{K}$&(Little) group&Irrep&$\mathcal{O}_{\pi\pi}$&$\mathcal{O}_\rho$ \\
\hline
48&$(0,0,0)$&$O_{h}$&$T_1$&$\pi(1,0,0)\pi(-1,0,0)$&$\rho_x$ \\
48&$(0,0,1)$&$C_{4v}$&$E$&$\pi(1,0,0)\pi(-1,0,1) - \pi(-1,0,0)\pi(1,0,1)$&$\rho_x$ \\
48&$(0,1,1)$&$C_{2v}$&$A_1$&$\pi(1,0,0)\pi(-1,1,1)+\pi(-1,0,0)\pi(1,1,1)$&$\rho_y+\rho_z$\\
48&$(0,1,1)$&$C_{2v}$&$B_1$&$\pi(1,0,0)\pi(-1,1,1)-\pi(-1,0,0)\pi(1,1,1)$&$\rho_x$\\
64&$(0,0,1)$&$C_{4v}$&$E$&$\pi(1,0,0)\pi(-1,0,1) - \pi(-1,0,0)\pi(1,0,1)$&$\rho_x$ \\
64&$(0,1,1)$&$C_{2v}$&$A_1$&$\pi(1,0,0)\pi(-1,1,1)+\pi(-1,0,0)\pi(1,1,1)$&$\rho_y+\rho_z$\\
64&$(0,1,1)$&$C_{2v}$&$B_1$&$\pi(1,0,0)\pi(-1,1,1)-\pi(-1,0,0)\pi(1,1,1)$&$\rho_x$\\
\hline
$K\pi$&&&& \\
$N_s$&$\mathbf{K}$&(Little) group&Irrep&$\mathcal{O}_{K\pi}$&$\mathcal{O}_{K^*}$ \\
\hline
48&$(1,1,0)$&$C_{2v}$&$B_2$&$\pi(1,0,0)K(0,1,0)$&$K^*_x - K^*_y$\\
64&$(0,0,0)$&$O_{h}$&$T_1$&$\pi(1,0,0)K(-1,0,0)$&$K^*_x$ \\
64&$(0,0,1)$&$C_{4v}$&$E$&$\pi(1,0,0)K(-1,0,1) - \pi(-1,0,0)K(1,0,1)$&$K^*_x$ \\
64&$(0,1,1)$&$C_{2v}$&$A_1$&$\pi(1,0,0)K(-1,1,1)+\pi(-1,0,0)K(1,1,1)$&$K^*_y+K^*_z$\\
64&$(0,1,1)$&$C_{2v}$&$B_1$&$\pi(1,0,0)K(-1,1,1)-\pi(-1,0,0)K(1,1,1)$&$K^*_x$\\
\end{tabular}
\end{ruledtabular}
\end{table*}

By using the two volumes and a number of moving frames, we are able to access several points within the regions of interest
around the expected positions of the $\rho$ and $K^*$ resonances.
The kinematic points we use are given in Table~\ref{tab:irreps},
where
\begin{equation}
\mathbf{K}=\frac{L}{2\pi}\mathbf{P}
\end{equation} denotes an
integer-valued lattice momentum vector.
The choice of momenta and representations is based on the requirement that
the non-interacting two-particle states lie within or close to the expected
resonance widths. To allow reuse of the generated
propagators, we restrict ourselves to $\mathbf{k}_1=\mathbf{p_1}L/(2\pi)=(1,0,0)$.
For each total momentum $\mathbf{P}$, we have to construct interpolators which transform according to a definite irreducible representation (irrep) of the little group of allowed cubic rotations once a Lorentz boost has been applied.
We construct the interpolators using the information about the little groups given in Ref.~\cite{Gockeler:2012yj}.
The irreps we work with and the (one- and two-particle) interpolators that transform according to each representation are also listed in Table~\ref{tab:irreps}.
We use Schoenflies notation (see, for example, Ref.~\cite{Hamermesh:100343}) for the names of the groups and irreps.

The necessary quark line contractions are depicted in
Fig.~\ref{fig:contract}, where the first row includes
two-particle to two-particle transitions and the second
row one- to one- as well as two- to one-meson transitions.
We use stochastic $\mathbb{Z}_2+i\mathbb{Z}_2$ wall sources
at one time slice for each spin component
and, for the contractions involving the two-particle interpolators, sequential inversions to generate all the contributing diagrams,
following Refs.~\cite{Aoki:2007rd,Aoki:2011yj}.
To compute the top left contraction of Fig.~\ref{fig:contract},
it is necessary to use two stochastic sources per configuration.
We use this minimum number of estimates per configuration
as the gauge noise dominates.
We further reduce the computational cost by fixing
$\mathbf{k}_1$ to $(1,0,0)$.
Even with this restriction, we can obtain several interesting levels around the expected positions of the $\rho$ and $K^*$ resonances.
Moreover, we only compute the full $\pi\pi \to \pi\pi$ correlator 
from $t = 6a$ to $t = 17a$, where we anticipate that on 
one hand the signal is only moderately
polluted by excited state contributions and on the other hand statistical
errors are still tolerable.
We are also able to ``recycle'' many propagators in both $\pi\pi$ and $K\pi$
scattering.

Adding this up,
in our implementation the total number of solves required on each
configuration is 
\begin{equation} N_{\mathrm{vec}} \left[N_{\mathrm{smear}}+N_{p_1} (1+18N_{p_2} + 3N_{\mathrm{times}} )\right]\,, \label{eq:cost} \end{equation}
where $N_{\mathrm{vec}}=8$ is the number of noise sources used
(four spin components times two different vectors),
$N_{\mathrm{smear}}=4$ is the number of one-particle smearing levels
(three plus one derivative source, see above),
$N_{p_1}=1$ and $N_{p_2}$ (see Table~\ref{tab:irreps}) are the numbers of momenta calculated and $N_{\mathrm{times}}=12$
($t=6a$ up to $t=17a$) is the number of time slices for which the box diagrams shown in the top middle and top right of Fig.~\ref{fig:contract} are calculated.
For the $N_s=48$ and $N_s=64$ lattices, evaluating the full
eight by eight matrices of correlators
for each moving frame amounts to inverting the
strange quark Wilson matrix 80 and 120 times,
respectively, and the light quark matrix 824 and 808 times.
Note that the number of solves
required to compute a ``traditional'' point-to-all propagator
is twelve, i.e.\ the present
scattering computation is by a
factor of about $40$ more expensive than a conventional
determination of the spectrum of stable light hadrons for one
quark smearing level (twelve strange and twelve light quark inversions
on each volume).

The momenta injected are not indicated in Fig.~\ref{fig:contract} and
the correlator is the sum of all allowed momentum projections;
some irreps require a combination of two
related pairs of momenta and, in $\pi\pi$ scattering, we can
interchange the momenta $\mathbf{p}_1$ and $\mathbf{p}_2$ carried
by each pion at the sink.
Similarly, we ensure that the one-particle to one-particle correlators ---
depicted in the lower left of the figure --- transform according
to the irreps given in Table~\ref{tab:irreps},
by taking the corresponding combinations of vector meson polarizations.
The contractions for $\pi\pi \to \rho$ and $\rho \to \pi\pi$
are complex conjugates and it is computationally cheaper to only
calculate one of them. (We do this for $\pi\pi\to\rho$.) 
For the remaining correlation matrix elements with $i\neq j$
(one- to one-particle), we
average over $C_{ij}$ and $C^*_{ji}$.

\subsection{Extraction of energy levels and phase shifts}
\label{sec:extra}
For each kinematic situation, we construct an eight times eight matrix of
correlators for our basis of interpolators in the way described above.
The element of this matrix
for a source interpolator $\mathcal{O}_j$ and a
sink interpolator $\mathcal{O}_i$ is given as
\begin{equation} C_{ij}(t) = \langle 0| \hat{\mathcal{O}}_i(t) \hat{\mathcal{O}}_j^\dag(0)  |0 \rangle\,. \label{eq:corrdef} \end{equation}
The spectral decomposition can be written as
\begin{equation}C_{ij}(t) = \sum_\alpha \frac{Z^i_\alpha Z^{j*}_\alpha}{2 E^\alpha} e^{-E^\alpha t}\,,\label{eq:spect}\end{equation}
where $Z_{\alpha}^i=\langle 0|\hat{\mathcal{O}}_i|\alpha\rangle$ is the overlap factor
of the state created by the operator
$\hat{\mathcal{O}}_i^{\dagger}$ with the physical
state $|\alpha\rangle$ of energy $E^{\alpha}$.
We extract the energy levels $E^\alpha$ by solving the generalized eigenvalue problem \cite{Michael:1985ne,Luscher:1990ck,Blossier:2009kd}
\begin{equation}C(t) u^{\alpha}(t) = \lambda^{\alpha}(t_0,t) C(t_0) u^{\alpha}(t)\,,
\label{eq:gevp}
\end{equation}
where the energy levels
can be obtained from the dependence
$\lambda^{\alpha}(t_0,t) \sim e^{-E^{\alpha}(t-t_0)}$ at large times.

The energies we extract are in the lab frame, so we denote these as $E_L$.
The phase shift, however, is extracted in the centre of momentum
frame, i.e.\ in the rest frame of the $\pi\pi$- or $\pi K$-system. 
It is straightforward to convert the lab frame energies
$E_L$ into the corresponding centre of momentum
frame energies $E_{cm}$.

The lab frame energy of the two-meson state
is given as
\begin{equation} E_L= \sqrt{\mathbf{p}_1^2 + m_1^2} + \sqrt{\mathbf{p}_2^2 + m_2^2}\,, \end{equation}
where the $m_i$ are the pion (or kaon) masses and the $\mathbf{p}_i$ their momenta. In the absence of interactions the $\mathbf{p}_i^2$ are integer multiples of $(2\pi/L)^2$. The invariant squared  energy in the centre of momentum frame is
\begin{equation}  
E_{cm}^2 = E_L^2 - \mathbf{P}^2\,,
\end{equation}
where $\mathbf{P}$ is the total momentum of the $\pi\pi$ (or the
$K\pi$) system. The square of the
momentum of each of the pseudoscalars in the centre of momentum frame
is given by
\begin{equation} p_{cm}^2 = \frac{\left(E_{cm}^2 - (m_1+ m_2)^2\right)\left(E_{cm}^2 - (m_1- m_2)^2\right)}{4 E_{cm}^2}\,. 
\label{eq:kcm} 
\end{equation}

The phase shift is extracted, comparing
the centre of momentum frame spectrum to the energy levels
allowed by the residual cubic symmetry (little group) that corresponds
to the boost applied.
For each irrep, this involves an expression in terms of
generalized zeta functions,
derived in Refs.~\cite{Leskovec:2012gb,Gockeler:2012yj}.
For the numerical calculation of these functions, we use the representation
given in Ref.~\cite{Gockeler:2012yj}.

The generalized zeta function is a function of the real-valued variable
$q = p_{cm} L/(2\pi)$:
\begin{equation}
Z_{\ell m}(q^2) = \sum_{\mathbf{z}} \frac{\mathcal{Y}_{\ell m}(\mathbf{z})}{\mathbf{z}^2-q^2}\,, \label{eq:zeta} \end{equation}
where
$\mathcal{Y}_{\ell m}(\mathbf{z}) = |\mathbf{z}|^{\ell} Y_{\ell m}(\mathbf{e}_z)$ with $\mathbf{e}_z = \mathbf{z}/|\mathbf{z}|$ and $Y_{\ell m}$ are
the usual spherical harmonics.
The sum is over $\mathbf{z}$, the allowed momentum vectors in the boosted
frame, see, e.g., Ref.~\cite{Gockeler:2012yj}.

For each irrep we have to consider mixing between different continuum
partial waves. The relevant determinants from which the phase shifts
can be extracted are listed in Ref.~\cite{Gockeler:2012yj}.
Here, we neglect possible mixing with partial waves
$\ell\neq 1$. The $s$-wave can only contribute
to $K\pi$ scattering. Moreover, mixing of $\ell=0$ into $\ell=1$ is only allowed
for the $\mathbf{K}=(0,1,1)$ $A_1$ irrep. We will address
this case in Sec.~\ref{sec:consistent} below. Since the $\pi\pi$ and
$K\pi$ interactions have a finite range, contributions of
higher partial waves are suppressed. The $\ell=3$ $\pi\pi$ phase
shift was determined recently by Wilson and
collaborators~\cite{Wilson:2015dqa} at $m_{\pi}\approx 236\,\textmd{MeV}$
who indeed found $\delta_{3}\approx 0$ near the resonance,
within small errors. We conclude that limiting ourselves to
$\ell\leq 1$ appears reasonable.

Subsequently, we parameterize the phase shift as a function of the centre of
momentum frame energy using a Breit-Wigner (BW) ansatz: 
\begin{equation} \tan\delta = \frac{g^2}{6\pi} \frac{p_{cm}^3}{E_{cm}(m_R^2-E_{cm}^2)}\,. \label{eq:bw} \end{equation}
From this parametrization,\footnote{We consider
alternative parametrizations in Sec.~\ref{sec:consistent}.} we can extract the mass of
the resonance $m_R$ and its width can be found from the coupling $g$ as
\begin{equation}
\label{eq:decay}
\Gamma = \frac{g^2}{6\pi} \frac{p_R^3}{m_R^2}\,,
\end{equation}
where $p_R$ is the momentum carried by each particle in the centre
of momentum frame at $\delta=\pi/2$, i.e.\ $p_R$ is given by
$p_{cm}$ of Eq.~\eqref{eq:kcm} for $E_{cm}=m_R$.

\section{Results}
\label{sec:results}
\subsection{Determination of the energy levels}
Following the generalized eigenvalue procedure detailed
in Sec.~\ref{sec:extra} above, we separately analyse
the eight by eight matrices that cross-correlate states created
by one- and two-particle interpolators for the seven $\pi\pi$
and five $K\pi$ channels listed in Table~\ref{tab:irreps}, and
obtain the respective ground and first excited state energies.
We are able to resolve these energies most easily using
sub-matrices of correlators containing only three interpolators ---
one of which always is the two-particle interpolator $\mathcal{O}_{\pi\pi}$ or
$\mathcal{O}_{K\pi}$ of Table~\ref{tab:irreps}. The
single-particle interpolators used in the final
analysis are only of the type $\bar{\psi} \gamma_j \psi$.
However, we have checked these results
against employing other sub-matrices and found consistency
of the effective masses, but no improvement.
The results turned out very similar but often noisier
when replacing one $\bar{\psi} \gamma_j \psi$ interpolator by
$\bar{\psi} \gamma_j \gamma_t \psi$ while
the $\bar{\psi} \nabla_j \psi$ interpolator increased
the statistical errors very significantly,
in particular for states with total momentum
$\mathbf{K} = (0,1,1)$.

To save computer time we only evaluated the box diagrams in the
top middle and top right
of Fig.~\ref{fig:contract} for $17a\ge t\ge  6a$. The top left
diagram contains two traces and naively increases like $L^6$ while the
quark-line connected box diagrams have magnitudes $\propto L^3$.
Due to this relative suppression,
these can only become important at times of at least a similar
magnitude as
the inverse energy gap between $I=2$ and $I=1$ $\pi\pi$ (or
$I=3/2$ and $I=1/2$ $K\pi$) states and probably their
contribution to the $\pi\pi\to\pi\pi$ and $K\pi\to K\pi$
entries can be neglected at
$t<6a$. Nevertheless, to be on the safe side,
in our generalized eigenvector analysis
we set $t_0=6a\approx 0.43\,\textmd{fm}$. 

\begin{figure}
\includegraphics[width=0.48\textwidth]{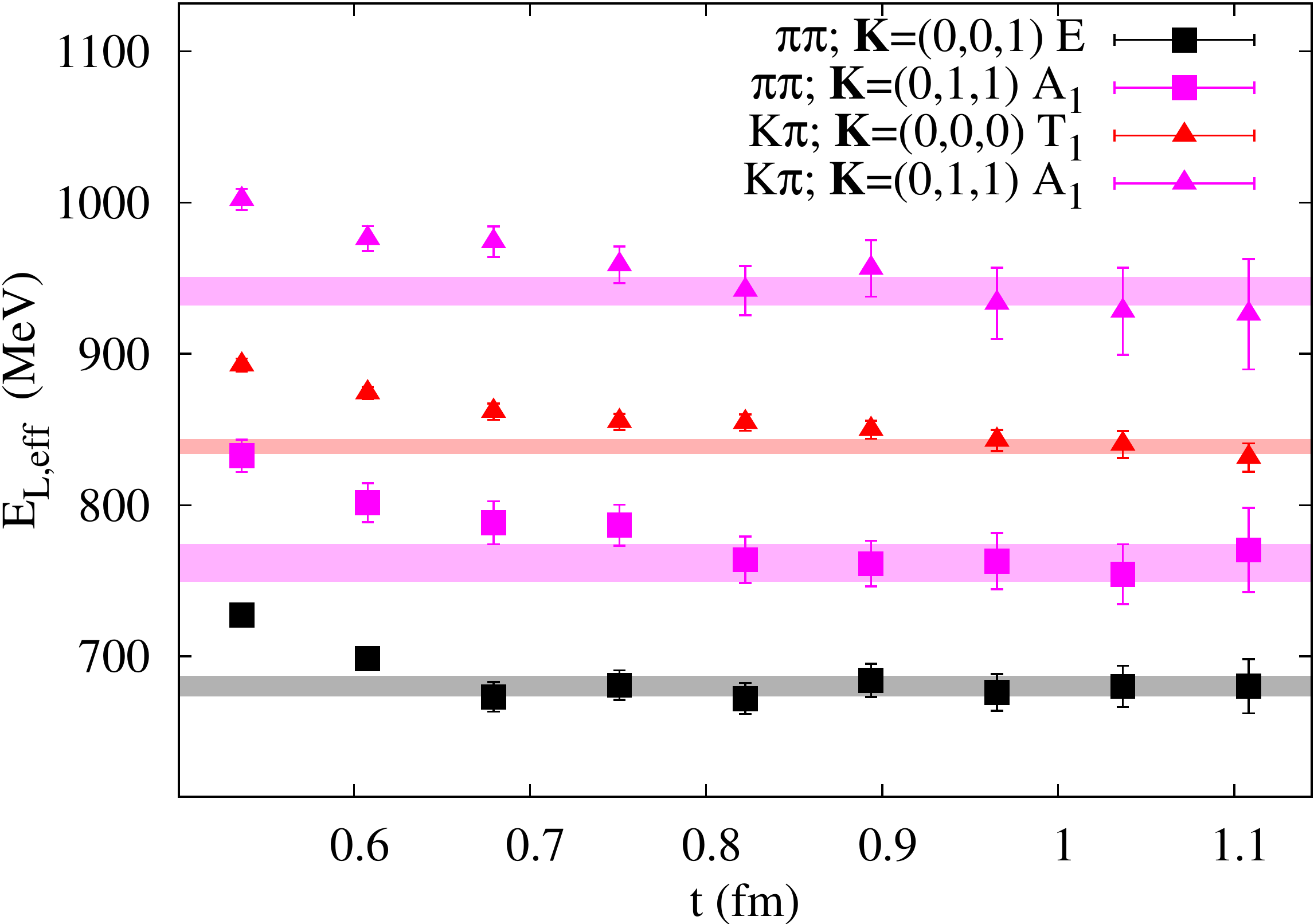}
\caption{Effective masses for some $\pi\pi$ and $K\pi$
channels. The error bands correspond to the fit results.}
\label{fig:meff}
\end{figure}

\begin{figure*}[h]
\includegraphics[width=.98\textwidth]{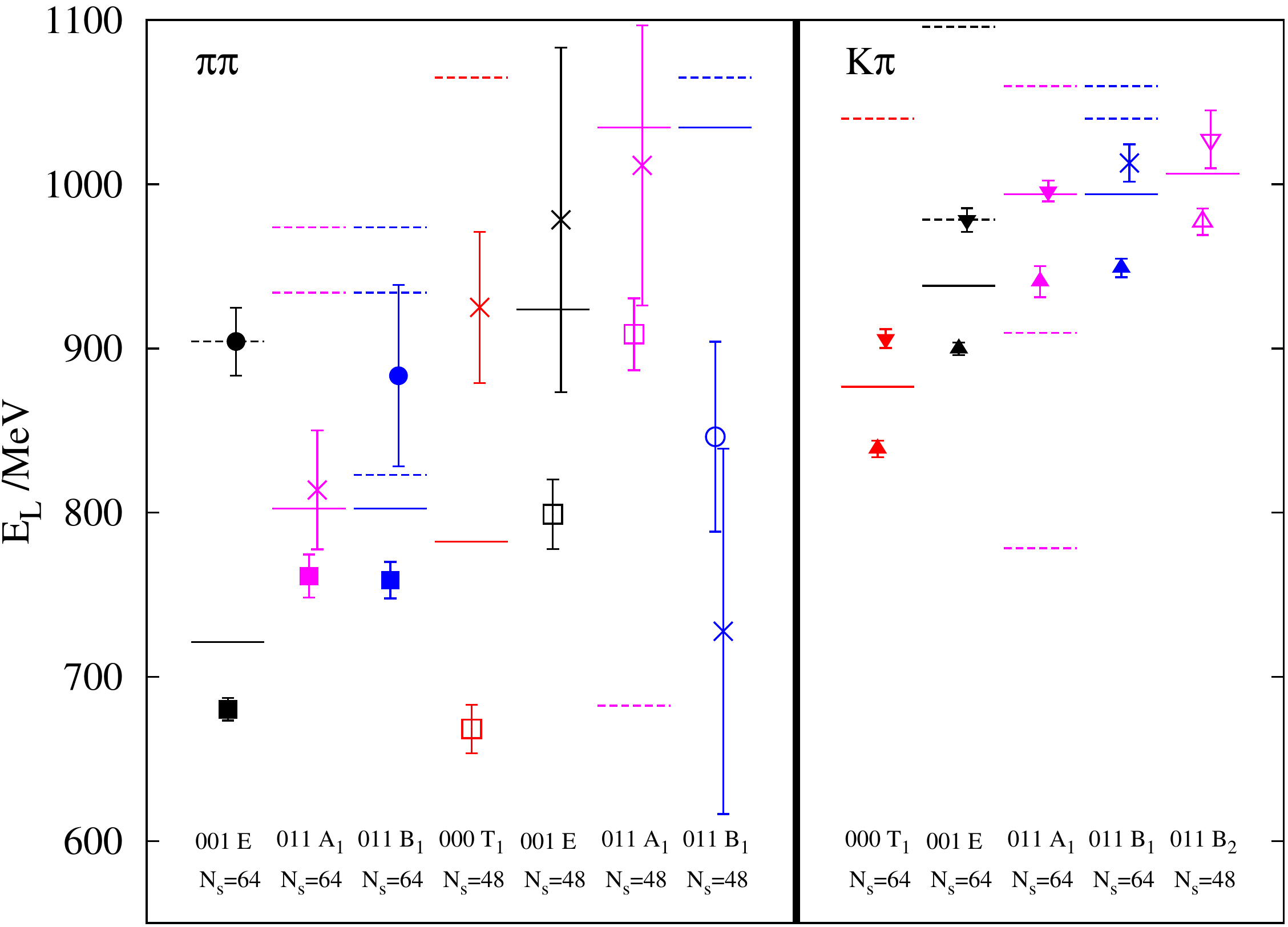}
\caption{Energy levels of the $\rho|\pi\pi$ (left) and $K^*|K\pi$ (right)
systems in finite boxes of linear sizes $N_sa=48a\approx 2.6/m_{\pi}^{\infty}\approx 3.4\,\textmd{fm}$ and
$N_sa=64a\approx 3.5/m_{\pi}^{\infty}\approx 4.6\,\textmd{fm}$
for different lattice momenta and representations in the laboratory
frame. Horizontal lines
correspond to the energy levels of a non-interacting two-particle system.
Squares and upward pointing triangles
indicate ground states, circles and downward pointing triangles
first excited states.
Open symbols correspond
to the smaller volume and full symbols to the larger volume.
Crosses are for levels that are not used in our subsequent phase shift analysis.
Note that for $\pi\pi$ scattering the excited states in both
$\mathbf{K}=(0,0,1)$ $E$ channels are above the respective non-interacting
$4\pi$ thresholds (not shown).}
\label{fig:spectra}
\end{figure*}

\begin{figure*}
\centering
\includegraphics[width=0.83\textwidth]{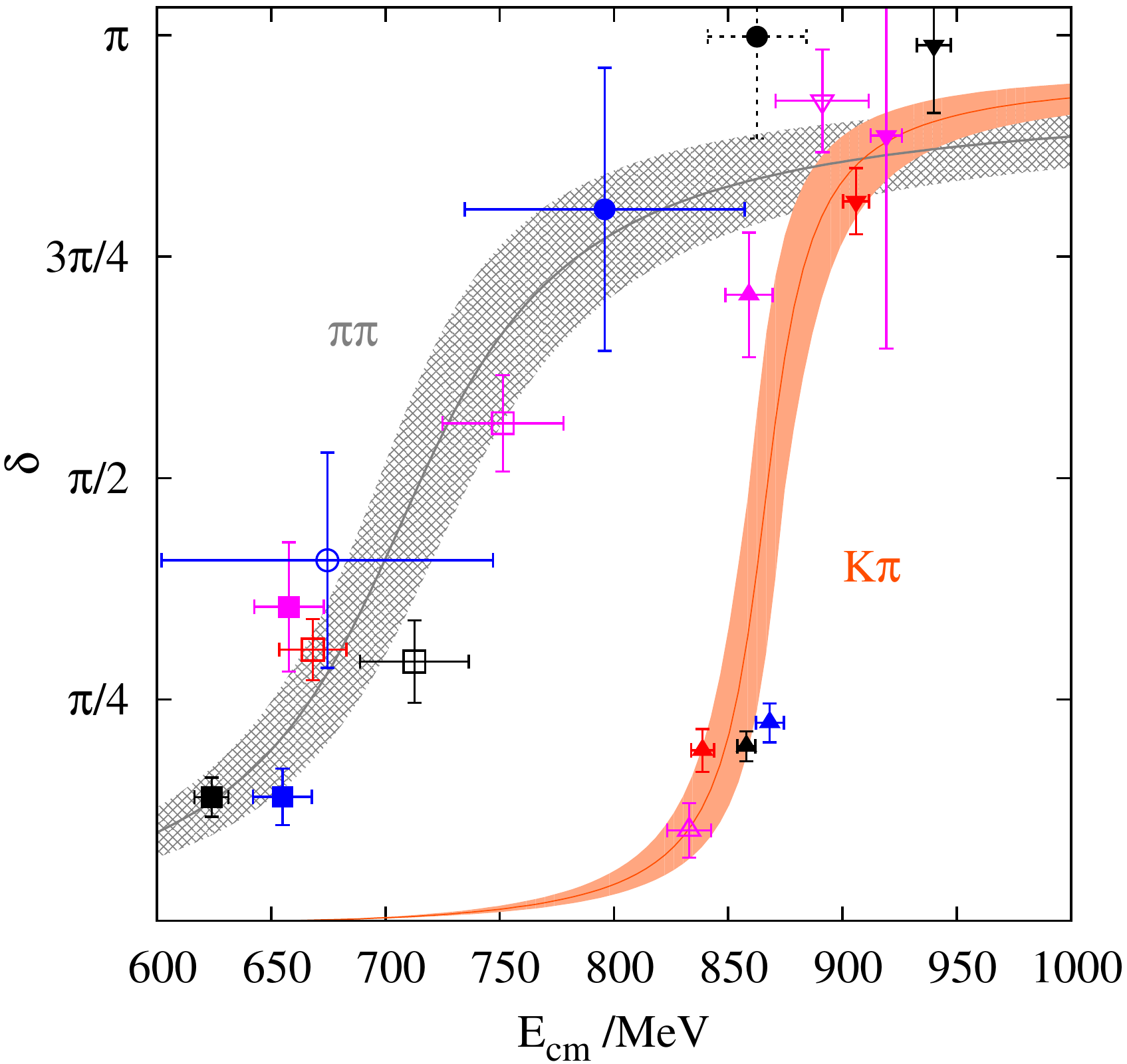}
\caption{The phase shift as a function of the centre of momentum frame
energy, $E_{cm}$, for $p$-wave $\pi\pi$ scattering around the $\rho$
resonance and $K\pi$ scattering around the $K^*$ resonance. The data
correspond to the lab frame energies shown in Fig.~\protect\ref{fig:spectra},
with matched colours and symbols. The curves with error bands
are Breit-Wigner parametrizations. The dashed error bar
indicates a point in $\pi\pi$ scattering which lies above
the four-pion threshold.}
\label{fig:phaseshift}
\end{figure*}

We show effective masses
\begin{equation}
E_{L,\mathrm{eff}}^{\alpha}(t+a/2)=\frac{1}{a}
\ln\frac{\lambda^{\alpha}(t_0=6a,t)}{\lambda^{\alpha}(t_0=6a,t+a)}
\end{equation}
for some of our $\pi\pi$ and $K\pi$ eigenvalues,
see Eq.~\eqref{eq:gevp},
in Fig.~\ref{fig:meff}, for the region
$t> t_0+a$. To enable better comparison to other studies,
we display the data in physical units.
The effective masses
are typically consistent with plateaus
between $t=10a\approx 0.71\,\textmd{fm}$ and $17a\approx
1.22\,\textmd{fm}$, which is our most frequent fit range, although
there are differences between the channels. The
$K\pi$ $T_1$ channel shown in the figure is an extreme example,
where the fit range starts at $t=14a\approx 1\,\textmd{fm}$.

Of particular interest are the
$\mathbf{K}=(0,1,1)$ $A_1$ channels. The
non-interacting ground states in this irrep correspond to a momentum
distribution $\mathbf{k}_1=\bf{0}$ and $\mathbf{k}_2=\mathbf{K}=(0,1,1)$
among the two pseudoscalar mesons that differs
from the one used in constructing our two-particle interpolators
($\mathbf{k}_1=(1,0,0)$ and $\mathbf{k}_2=\mathbf{K}-\mathbf{k}_1=(-1,1,1)$).
In principle, these correlation functions could decay towards
the lower lying states. However, we find no indication
for this in our data, see Fig.~\ref{fig:meff}, and conclude that
our interpolators effectively decouple from these energy levels.

The resulting lab frame energy levels $E_L$ are shown in
Fig.~\ref{fig:spectra} both for the $\pi\pi$ and $K\pi$ channels.
The scale is set using $a^{-1}=2.76\,\textmd{GeV}$, ignoring
the 3\% overall scale uncertainty for the moment being.
The statistical errors are obtained using the jackknife procedure.
Only two $\pi\pi$ levels are above the four-pion threshold
(the excited states in the $\mathbf{K}=(0,0,1)$ $E$ irrep), one of
which will be disregarded in any case in the phase shift analysis below.

In the figure, we also show the energies of the non-interacting two-particle
states.
The solid horizontal lines are the non-interacting levels corresponding
to the two-particle interpolators explicitly included in our basis
(given in Table~\ref{tab:irreps}), while the dashed lines correspond to
other distributions of the momentum among the non-interacting
pseudoscalar mesons. As
we have not included interpolators that explicitly resemble
these momentum configurations, we cannot rely on our extracted energy
levels to be sensitive to their presence and ignore these
non-interacting levels in our phase shift analysis.
As already discussed above, in the $A_1$ case the
non-interacting ground states are lower in energy
than the levels that correspond
to the momentum distribution we have implemented (solid lines).
Nevertheless, we see no evidence of any coupling of the interpolators
within our basis to these states, see Fig.~\ref{fig:meff}.
Note that for the $N_s=64$ $\pi\pi$ channel
this level lies at $561\,\textmd{MeV}$, below the energy region shown
in Fig.~\ref{fig:spectra}.

Levels that are irrelevant, due to large
statistical errors for the resulting phase shifts,
will be excluded from our subsequent analysis.
These levels are depicted as crosses in Fig.~\ref{fig:spectra}.
We remind the reader that the deviations of the measured energy levels
shown in the figure from the non-interacting two-particle levels
(solid lines) are due to the $\rho$ and $K^*$ resonances and encode
the resonance parameters.

\subsection{Phase shift and resonance parameters}
The centre of momentum frame energies $E_{cm}$ and
phase shifts $\delta(E_{cm})$ can both be extracted
from measured lab frame energy levels $E_L$ in a given irrep,
see Sec.~\ref{sec:extra}, where
we assume $m_{\pi}=149.5\,\textmd{MeV}$,
in spite of the fact that the measured pion mass on the small volume
is larger by $10\,\textmd{MeV}$. This will be addressed in
Sec.~\ref{sec:consistent}
below.

We plot $\delta(E_{cm})$ in Fig.~\ref{fig:phaseshift}, using the same
colour and symbol scheme as in Fig.~\ref{fig:spectra}.
As explained above, in our determination of the phase shift we assume
that one value of
$\ell$ ($\ell=1$) dominates, such that there is a one to one correspondence
between the extracted energy levels and the points in the phase shift curves.
For clarity we omit all data points from the figure with errors
on the phase shift in excess of $\pi/5$ (marked as crosses
in Fig.~\ref{fig:spectra}). These have little statistical
impact and will therefore be excluded from our analysis.

The $\pi\pi$ and $K\pi$ phase shifts are each fitted to the BW
resonance form given in Eq.~\eqref{eq:bw}.
Our fit to the $\pi\pi$ phase shift results in
$\chi^2 /\mathrm{d.o.f}  = 8.9/7$ and for the $K\pi$ phase shift
we obtain $\chi^2 / \mathrm{d.o.f.} = 19.2/7$.
These fits are included in Fig.~\ref{fig:phaseshift} (the grey hashed
band for $\pi\pi$ scattering and the solid orange one for $K\pi$ scattering).
In the $\pi\pi$ case the dashed data point of the figure
is slightly above the respective $4\pi$ threshold.
However, as discussed in the introduction,
the effect of this inelastic threshold is expected to be negligible.
Moreover, excluding this point from the fit only produces
a hardly visible change. Since we have exact isospin
symmetry in place, decays into three-pion final states are not possible.

Figures~\ref{fig:spectra} and \ref{fig:phaseshift} clearly show an increase
in statistical noise when going to smaller quark masses:
The $\pi\pi$ scattering data have considerably larger error bars
than the $K\pi$ data. From the BW fits shown, we find the values
\begin{align}
m_{\rho} = 716(21)(21)\,\textmd{MeV}\,,&\quad
m_{K^*} = 868(8)(26)\,\textmd{MeV}\,,\label{eq:mass}\\
g_{\rho\pi\pi} = 5.64\pm 0.87\,,&\quad
g_{K^*K\pi} = 4.79\pm 0.49\,,\label{eq:coupling}\\
\Gamma_{\rho} = 113(35)(3)\,\textmd{MeV}\,,&\quad
\Gamma_{K^*} = 30(6)(1)\,\textmd{MeV}\,,\label{eq:gamma}
\end{align}
for $\pi\pi$ and $K\pi$ scattering, where the first errors
are statistical and the second errors reflect our 3\% overall
scale uncertainty~\cite{Bali:2012qs}. 
In the last row we also quote the
corresponding decay widths, obtained via Eq.~\eqref{eq:decay}.
From a given parametrization of the $p$-wave phase shift, assuming
partial wave unitarity and ignoring further inelastic thresholds,
we can analytically continue to the second (unphysical) Riemann
sheet (see, e.g., Ref.~\cite{Burkhardt}) and determine
the position of the resonance pole. Using the
BW parametrization, for the $\rho$ and $K^*$ resonances
we find $\sqrt{s_R}=[707(17)-\I\,55(18)]\,\textmd{MeV}$
and $\sqrt{s_R}=[868(10)-\I\,14.4(3.4)]\,\textmd{MeV}$, respectively.
These numbers are consistent with $\sqrt{s_R}=m_R-\I\,\Gamma/2$
from the BW fits
Eqs.~\eqref{eq:mass} and \eqref{eq:gamma}.
Note, however, that $\textmd{Re}\sqrt{s_{\rho}}$ is by about half a
standard deviation smaller than the BW fit parameter $m_{\rho}$.
In Sec.~\ref{sec:consistent} we will explore in detail
the parametrization dependence of these results.

\begin{figure}
\includegraphics[width=0.48\textwidth]{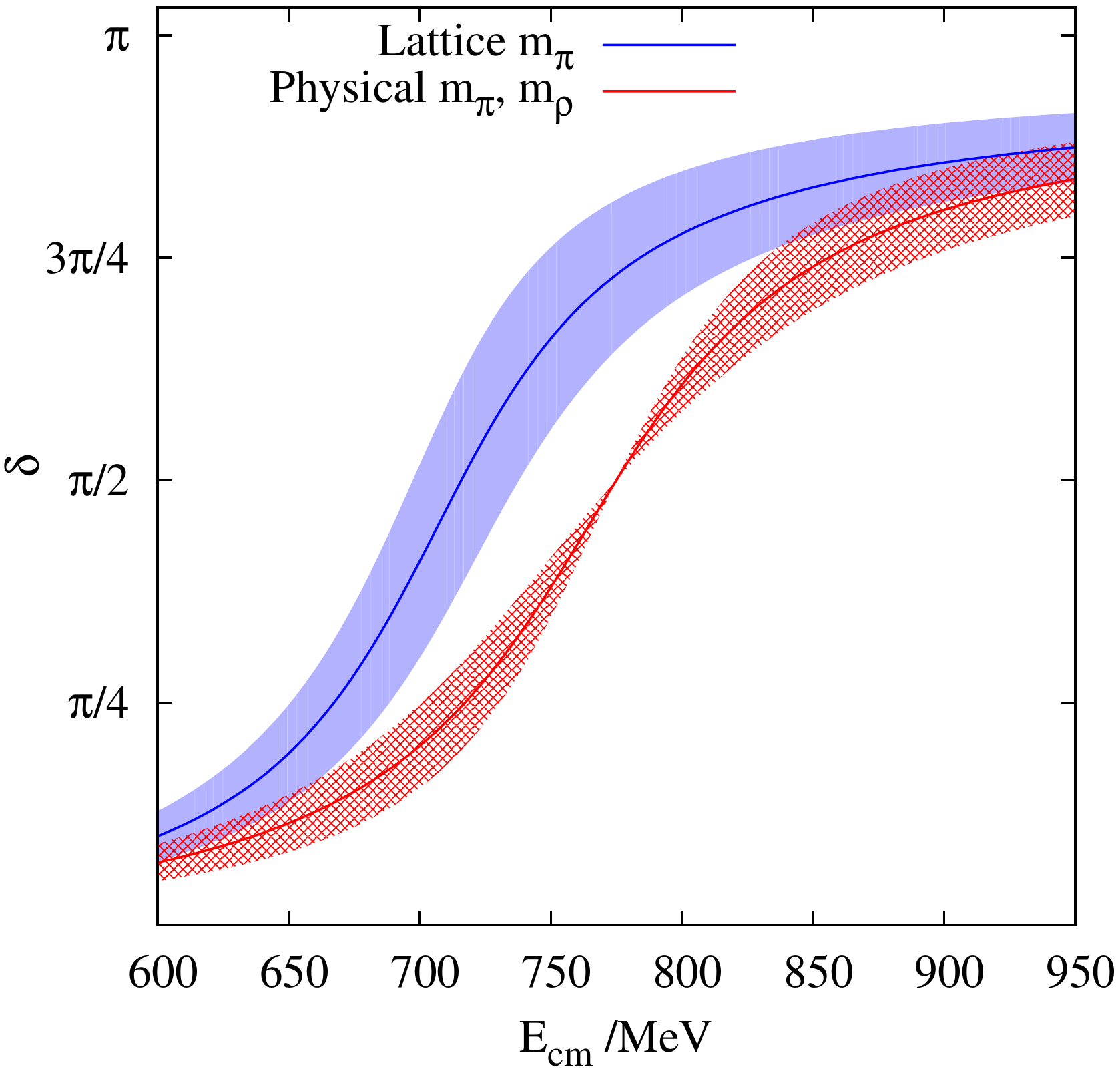}
\caption{We compare phase shift curves for $\pi\pi$ scattering around the $\rho$ resonance for our fitted Breit-Wigner resonance (solid blue band) and one with the fitted coupling but physical pion and $\rho$ masses (hashed red band).}
\label{fig:pipips}
\end{figure}

We emphasize that our study was carried out
at a single lattice spacing only, which is not reflected in
the errors given above.
Both resonant masses come out smaller
than the experimental values, $775\,\textmd{MeV}$ and
$896\,\textmd{MeV}$, respectively.
The reduced decay phase space, due to a 10\% heavier than physical
pion, in conjunction with somewhat smaller than physical
resonance masses,
is the main reason why our decay widths appear to be somewhat
below the experimental ones,
$\Gamma_{\rho} \approx 148\,\textmd{MeV}$ and
$\Gamma_{K^*} \approx 47\,\textmd{MeV}$, although
this difference is only statistically significant for the $K^*$.
The coupling $g_{\rho\pi\pi}$ is consistent with the experimental value
$g_{\rho\pi\pi}\approx 5.93$ while our $g_{K^*K\pi}$ is slightly lower than
$g_{K^*K\pi}\approx 5.39$. The ordering
$g_{\rho\pi\pi}>g_{K^*K\pi}$ is reproduced, albeit within large errors.

In Fig.~\ref{fig:pipips} the $\pi\pi$ phase shift curve fitted to our
data is compared to the same curve with the $\pi$ and $\rho$ masses set
to their physical values~\cite{pdg}, but the coupling $g_{\rho\pi\pi}$ taken
from our fit Eq.~\eqref{eq:coupling}. The latter curve is
forced to run through $\delta=\pi/2$ at the fixed resonant
mass $E_R=E_{cm}=775\,\textmd{MeV}$ as we use the BW parametrization.
Since our value of $g_{\rho\pi\pi}$ agrees with experiment, experimental
data will be described by the hashed red band.
Again, there is an overall scale setting uncertainty
of 3\% on $E_{cm}$, corresponding to $20\,\textmd{MeV}$,
that we do not display as well as other systematics,
most notably a 10\% heavier than physical pion and a fixed
lattice spacing. The figure illustrates that also in terms
of the width of the resonance we are close to the physical case.
Previous studies of $\pi\pi$ scattering have not directly addressed
the physical limit, although unitarized chiral perturbation theory has been used
in Ref.~\cite{Bolton:2015psa} to
extrapolate lattice data obtained at
$m_\pi \approx 236\,\textmd{MeV}$~\cite{Wilson:2015dqa} to the physical point.

\subsection{Investigation of possible biases}
\label{sec:consistent}
Here we investigate the effects
on the extracted resonance parameters,
of the finite volume pion mass shift, of the BW parametrization
we use to fit $\delta(E_{cm})$ and of the presence of
inelastic thresholds. We also address the possibility of
an $\ell=0$ pollution for the case of $K\pi$ scattering.

The pion mass enters the generalized zeta function, Eq.~\eqref{eq:zeta},
via the calculation of the momentum carried by the two particles
in the centre of momentum frame, given by Eq.~\eqref{eq:kcm}.
We prefer to use the infinite volume pion and kaon masses
throughout because we are relating the spectra to scattering amplitudes
in an infinite volume.
For the larger $L=64a$ lattice size, the pion mass determined in the
finite volume and that extrapolated to infinite volume differ by as little
as $0.2\,\textmd{MeV}$~\cite{Bali:2014gha}.
However, the pion mass measured on the $L=48a$ configurations differs
from the infinite volume mass by $10\,\textmd{MeV}$ and the kaon mass
by $3\,\textmd{MeV}$.
Since pion exchanges around the
boundaries of the periodic box go beyond the L\"uscher formalism,
we have repeated the analysis using finite volume pion masses instead,
to explore these systematics.
For the $L=64a$ data the effect obviously is insignificant.
For $L=48a$ the phase shifts for the corresponding six points
(four for the $\rho$ resonance and two for $K^*$) depicted in
Fig.~\ref{fig:phaseshift} (open symbols)
increase by values ranging from $0.03$ to $0.05$.
These differences
are considerably smaller than our errors on $\delta$.
Indeed, using these numbers instead, we find the $\rho$
and $K^*$ resonance parameters
$m_\rho = 713(18)\,\textmd{MeV}$,
$m_{K^*}=867(7)\,\textmd{MeV}$, $g_{\rho\pi\pi}=5.56(85)$
and $g_{K^*K\pi}=4.81(51)$,
in almost perfect agreement with
our main analysis employing the infinite volume pion mass
Eqs.~\eqref{eq:mass} and \eqref{eq:coupling}.
For instance, the central values for the masses
deviate by only $-3\,\textmd{MeV}$ and $-1\,\textmd{MeV}$, respectively.
Adding these systematics to the statistical errors in quadrature
has no impact.

\begin{table*}
\caption{
$\rho$ resonance: fit results for various phase shift models.
The square root of the resonance pole position $s_R$ may be used to
define $\sqrt{s_R}= m_R-\frac{\I}{2}\,\Gamma_R$. The errors given are
statistical only.}
\label{tab:rho}
\centering
\begin{ruledtabular}
\begin{tabular}{lccccc}
Model&$m_{\rho}/\textmd{MeV}$&$g_{\rho\pi\pi}$&other fit parameters&$\sqrt{s_R}/\textmd{MeV}$\\\hline
0: Eq.~\eqref{eq:bw} (BW)&716(21)&5.64(87)&---&$707(17) -\frac{\I}{2}\,110(36)$\\
1: Eq.~\eqref{eq:lin}    &717(23)&5.38(84)&$R=3(6)\,\textmd{GeV}^{-1}$&$714(26) -\frac{\I}{2}\,104(35)$\\
2: Eq.~\eqref{eq:exp}    &718(23)&5.34(84)&$\beta=0.16(15)\,\textmd{GeV}$&$716(29)-\frac{\I}{2}\,103(35)$\\
3: Eq.~\eqref{eq:com}    &717(23)&---&$B_0=1.31(45)$, $B_1=1.6(3.0)$&$714(26)-\frac{\I}{2}\,103(37)$
\end{tabular}
\end{ruledtabular}
\end{table*}
\begin{table*}
\caption{
$K^*$ resonance: fit results for various phase shift models, including and
excluding the two $\mathbf{K}=(0,1,1)$ $A_1$ irrep points that may also couple to
the $\ell=0$ partial wave. The errors given are statistical only.}
\label{tab:kstar}
\centering
\begin{ruledtabular}
\begin{tabular}{lcccccc}
Model&$A_1$ included&$m_{K^*}/\textmd{MeV}$&$g_{K^*K\pi}$&other fit parameters&$\sqrt{s_R}/\textmd{MeV}$\\\hline
0: Eq.~\eqref{eq:bw} (BW)&yes&868(8)&4.79(49)&---&$866(7) -\frac{\I}{2}\,30(7)$\\
1: Eq.~\eqref{eq:lin}    &yes&868(9)&4.78(44)&$R=6(29)\,\textmd{GeV}^{-1}$&$868(9) -\frac{\I}{2}\,30(7)$\\
2: Eq.~\eqref{eq:exp}    &yes&868(9)&4.80(47)&$\beta=0.13(45)\,\textmd{GeV}$&$867(10)-\frac{\I}{2}\,30(8)$\\
3: Eq.~\eqref{eq:com}    &yes&868(9)&---&$B_0=3.2(5.3)$, $B_1=8.2(28.9)$&$868(10)-\frac{\I}{2}\,29(7)$\\
0: Eq.~\eqref{eq:bw} (BW)& no&873(9)&5.08(43)&---&$871(8) -\frac{\I}{2}\,35(7)$\\
1: Eq.~\eqref{eq:lin}    & no&878(10)&5.09(38)&$R=1.2(1.6)\,\textmd{MeV}^{-1}$&$877(10) -\frac{\I}{2}\,36(6)$\\
2: Eq.~\eqref{eq:exp}    & no&887(7)&4.42(69)&$\beta=58(13)\,\textmd{MeV}$&$890(7)-\frac{\I}{2}\,27(9)$\\
3: Eq.~\eqref{eq:com}    & no&886(8)&---&$B_0=10(5)$, $B_1=44(25)$&$888(9)-\frac{\I}{2}\,24(6)$
\end{tabular}
\end{ruledtabular}
\end{table*}
Next, we replace the BW parametrization of the
scattering phase shift, see Eq.~\eqref{eq:bw}, by
other functional forms suggested in Ref.~\cite{Dudek:2012xn} and
references therein.
We write,
\begin{equation}
\tan\delta = \frac{E_{cm}\Gamma(E_{cm})}{m_R^2-E_{cm}^2}
\,,\quad
\Gamma^{(0)}(E_{cm}) = \frac{g^2}{6\pi} \frac{p_{cm}^3}{E_{cm}^2}\,,
\label{eqcite}
\end{equation}
where $\Gamma=\Gamma(m_R)$ is the resonance width
and the energy dependent width function $\Gamma(E_{cm})$ equals
$\Gamma^{(0)}(E_{cm})$ in the BW case.
In addition, we
use~\cite{VonHippel:1972fg,Li:1994ys,Pelaez:2004vs}\footnote{Note that $B_0$
is defined differently in Ref.~\cite{Dudek:2012xn} than here~\cite{Pelaez:2004vs}.}
\begin{align}\label{eq:lin}
\Gamma^{(1)}(E_{cm}) &= \frac{g^2}{6\pi} \frac{p_{cm}^3}{E_{cm}^2}\frac{1+(p_RR)^2}{1+(p_{cm}R)^2}\,,\\
\Gamma^{(2)}(E_{cm}) &= \frac{g^2}{6\pi} \frac{p_{cm}^3}{E_{cm}^2}\exp\left(\frac{p^2_{R}-p^2_{cm}}{6\beta^2}\right)\,,\label{eq:exp}\\
\label{eq:com}
\Gamma^{(3)}(E_{cm})&
=2\frac{p_{cm}^3}{E_{cm}^2}\\\nonumber
\times&\quad\left(B_0+B_1\frac{E_{cm}-\sqrt{s_0-E_{cm}^2}}{E_{cm}+\sqrt{s_0-E_{cm}^2}}\right)^{\!\!-1}\!\!\!\!,
\end{align}
where $s_0=(2m_{\pi}+m_R)^2$. The BW fit function
depends on two fit parameters, the resonant mass
$m_R$ and the coupling $g$, while the other parametrizations
depend on three parameters: $\Gamma^{(1)}$ contains the additional
parameter $R$, $\Gamma^{(2)}$ contains $\beta\sim 1/(\sqrt{6}R)$
and $g$ is replaced by $B_0$ and $B_1$ within $\Gamma^{(3)}$.

Our fit results for $\pi\pi$ scattering are shown in Table~\ref{tab:rho}.
In all cases the additional parameter ($R$, $\beta$ and $B_1$)
turned out to be consistent with zero. All the resonant masses we obtain are
in perfect agreement with the BW result shown in the first row. Also the widths are
compatible with the BW width $\Gamma^{(0)}=113(35)\,\textmd{MeV}$ of Eq.~\eqref{eq:gamma}
and the parameter $B_0=1.31(45)$ is consistent
with the expectation $B_0\approx 1.07$,
extracted from experimental data in Ref.~\cite{Pelaez:2004vs}.
Interestingly, we observe the numerically biggest difference
(half a standard deviation) between the energy at a phase shift
$\delta=\pi/2$,
$m_{\rho}=E_{cm}(\pi/2)$, and the real part of $\sqrt{s_R}$ for the BW parametrization.
We conclude from Table~\ref{tab:rho}
that within our precision, we can neither differentiate
between the different models nor distinguish
the pole position in the second Riemann sheet from the naively fitted
mass and width.

In our determination of the $\pi\pi$ energy levels, we noted that there was
one data point above the four-pion threshold (the dashed point of
Fig.~\ref{fig:phaseshift}).
Excluding this from any of our four fits, however, had
no impact worthy of mentioning.

For $K\pi$ scattering, in the case of the $\mathbf{K}=(0,1,1)$ $A_1$ irrep,
we cannot exclude the possibility of
a $\ell=0$ partial wave admixture. Therefore, we perform all
fits (setting $s_0=(m_{\pi}+m_{K}+m_R)^2$ in Eq.~\eqref{eq:com})
including and excluding the corresponding two data points, see the pink
solid triangles in Figs.~\ref{fig:spectra}
and \ref{fig:phaseshift}.
The resulting fit parameters and the position of the $K^*$ pole
are displayed in Table~\ref{tab:kstar}. When including the two $A_1$ points,
there is no sensitivity to the additional fit parameters and all the
results are remarkably stable.
Including and excluding these points, real and $2\I$ times the imaginary part
of $\sqrt{s_R}$ perfectly agree with the fitted masses and widths
obtained through Eqs.~\eqref{eqcite}--\eqref{eq:com}, as one would
expect for $\Gamma_{K^*}/m_{K^*}\approx 0.035\ll 1$.
Removing the two points, however, appears to increase the resonant mass.
Also the fit results become less stable 
since the BW fit has only five remaining degrees of
freedom while the other three fits have only four.

In conclusion, while we find $g_{K^*K\pi}$ to be very stable
against variations of the parametrization and of the number of points fitted, the
$K^*$ mass is somewhat affected by the latter. Therefore, we
allow for another systematic error of $10\,\textmd{MeV}$ to be added to the statistical
error shown in Eq.~\eqref{eq:mass} in quadrature:
\begin{equation}\label{eq:massk}
m_{K^*}=868(13)(26)\,\textmd{MeV}\,.
\end{equation}
\subsection{Investigation of an alternative method}
It is possible to estimate the value of the coupling $g_{\rho\pi\pi}$
directly from the
correlators, using the McNeile-Michael-Pennanen (MMP) method introduced in
Refs.~\cite{McNeile:2002az,McNeile:2002fh}
(also see Refs.~\cite{Gottlieb:1983rh,Gottlieb:1985rc}
for earlier, related work), if the momentum and volume are
selected such that the $\pi\pi$ energy is close to the resonant mass
$m_{\rho}=m_R$.
This method was also employed recently for studying the $\Delta$
resonance~\cite{Alexandrou:2013ata}.

Using the correlators defined in Eq.~\eqref{eq:corrdef},
with $\mathcal{O}_1$ and $\mathcal{O}_2$ being two- and one-particle
interpolators, we can extract (approximate) ground state energies
$E_{\pi\pi}$ and $E_{\rho}$ from $C_{11}(t)$ and $C_{22}(t)$ alone, respectively,
at times sufficiently small to avoid the higher level to decay into
the lower level (if $E_{\rho}\neq E_{\pi\pi}$) and large enough for
excited state contributions to be negligible.
In this situation, the ground state contribution to $C_{12}(t)$ reads
\begin{equation}
C_{12}(t) \approx xa\sum_{t'}\frac{Z_{\pi\pi}^1 Z_{\rho}^{2*}}{2\sqrt{E_{\pi\pi}E_{\rho}}} e^{-E_{\pi\pi}(t-t')}e^{-E_{\rho}t'} \,, \label{eq:c12} \end{equation}
where $Z_{\alpha}^i$ are the amplitudes to create the states $|\alpha\rangle$
using $\hat{\mathcal{{O}}}^{\dagger}_i$.
These overlap factors also appear
within $C_{11}(t)$ and $C_{22}(t)$ [see Eq.~\eqref{eq:spect}]
and will cancel as we are going to divide $C_{12}$ by an appropriate
combination of these two elements in Eqs.~\eqref{eq:divide} and
\eqref{eq:rratio} below.
The $\rho$ state created at $t=0$ will propagate to a
time $t'<t$, where it undergoes a transition into
$\pi\pi$. $x$ is the associated
$\rho\to\pi\pi$ transition amplitude and in Eq.~\eqref{eq:c12} we summed
over all possible intermediate times $t'$. The underlying assumption is that
the overlaps of $\hat{\mathcal{O}}_2^{\dagger}|0\rangle$ with $|\pi\pi\rangle$
and of $\hat{\mathcal{O}}_1^{\dagger}|0\rangle$ with $|\rho\rangle$ are
small and can be treated as perturbations, at least if
$t$ is not taken too large. Obviously, there are corrections of
higher order in $x$ to Eq.~(\ref{eq:c12}).

The coupling $g_{\rho\pi\pi}$ can then be estimated
from $x$ through~\cite{McNeile:2002fh}
\begin{equation}
\label{eq:g2}
g^2_{\rho\pi\pi} \approx \frac{L^3E_{cm}^3}{4p_{cm}^2}|x|^2\,.
\end{equation}
This can be seen as follows~\cite{McNeile:2002fh}. Fermi's Golden Rule relates
the decay width to the matrix element $x$ in the centre of
momentum frame: $\Gamma\approx |x|^2L^3p_{cm}E_{cm}/(24\pi)$. This can be
re-expressed in terms of $g^2$ through
$\Gamma=g^2p_{cm}^3/(6\pi E_{cm}^2)$, see Eq.~\eqref{eq:decay},
where $E_{cm}$ is taken at the
point $\delta=\pi/2$. The prefactor
$L^3p_{cm}E_{cm}/(24\pi)$ above contains the following contributions:
$2\pi$ from the Golden Rule, $L^3p_{cm}E_{cm}/(8\pi^2)$ from the
density of states, $1/2$ for a decay into identical
pions and $1/3$, averaging over one pion momentum direction for the
fixed $\rho$ polarization and momentum.

In Eq.~\eqref{eq:g2} several assumptions have been made:
(1) The Golden Rule is applicable, i.e.\ the $\pi\pi$ contribution
to the initial $\rho$ meson state is insubstantial and
the matrix element is not too large: $|x|t\ll 1$. This is
synonymous with neglecting terms of higher order
in $x$.
(2) The volumes are sufficiently large for continuous
density of states methods to be applicable.
(3) The $\pi\pi$ and $\rho$ states have a similar energy and,
in the centre of momentum frame, this is close to the resonant mass.
(4) $x$ does not change substantially when transforming it from
the lab to the centre of momentum frame.

\begin{table}
\caption{Estimates of $x$ and $g_{\rho\pi\pi}$ using the MMP
method~\protect\cite{McNeile:2002az,McNeile:2002fh}.
The entries are sorted in terms of a descending gap
$\Delta E= E_{\pi\pi}-E_{\rho}$. In the last row
we show our result Eq.~\protect\eqref{eq:coupling}
from the L\"uscher-type scattering analysis for comparison.}
\label{tab:gestimates}
\centering
\begin{ruledtabular}
\begin{tabular}{cccrrc}
$N_s$&$\mathbf{K}$&Irrep&$\Delta E/\textmd{MeV}$&$x/\textmd{MeV}$&$g_{\rho\pi\pi}$\\\hline
$48$&$(0,1,1)$&$A_1$& 135& 81(5)&5.54(30)\\
$48$&$(0,1,1)$&$B_1$&  95&106(7)&7.07(44)\\
$48$&$(0,0,1)$&$E$  &  16&124(6)&8.37(39)\\
$48$&$(0,0,0)$&$T_1$& -35&113(4)&7.54(28)\\
$64$&$(0,1,1)$&$A_1$&-122& 51(2)&5.19(17)\\
$64$&$(0,1,1)$&$B_1$&-140& 73(3)&8.18(22)\\
$64$&$(0,0,1)$&$E$  &-173& 81(2)&7.46(25)\\
\multicolumn{3}{l}{Full scattering analysis}&\multicolumn{1}{c}{---}&
\multicolumn{1}{c}{---}&5.64(87)
\end{tabular}
\end{ruledtabular}
\end{table}
\begin{figure}
\includegraphics[width=0.48\textwidth]{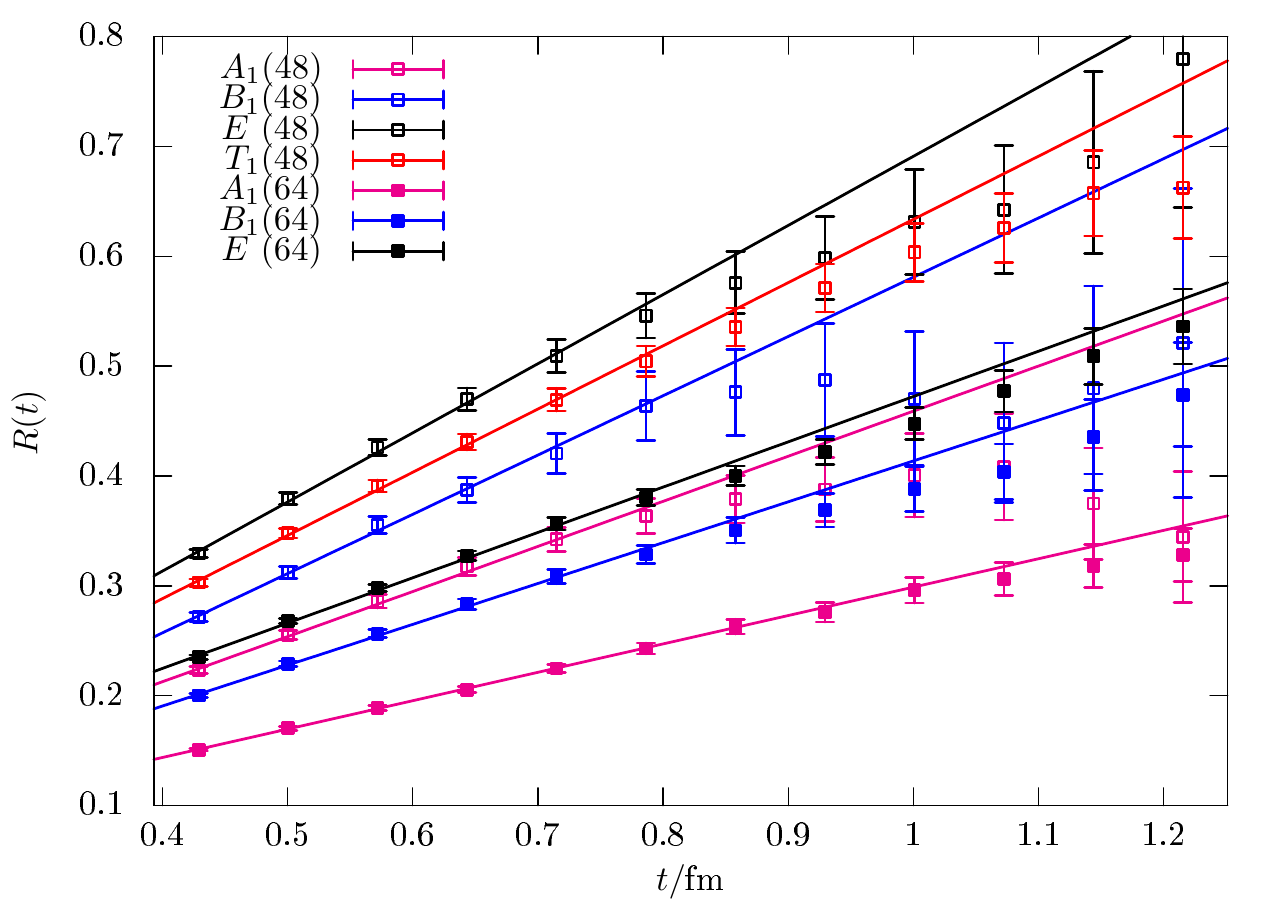}
\caption{The ratio of correlators $R(t)$
defined in Eq.~\eqref{eq:rratio}
for different irreps on the two volumes,
see Table~\protect\ref{tab:gestimates}, together
with linear fits to the first seven data points shown.}
\label{fig:directg}
\end{figure}

In the limit $E_{\pi\pi} = E_{\rho}$, summing over the intermediate
time $t'$, the ground state contribution to
Eq.~\eqref{eq:c12} has the time dependence $t \,e^{-E_{\pi\pi}t}$, while
excited states are suppressed by a power of $t$, relative to this.
In this case, $x$ can be found from a ratio of correlators as
\begin{equation}
\label{eq:divide} \frac{|C_{12}(t)|}{\sqrt{C_{11}(t)C_{22}(t)}} \approx \textmd{const.} +xt\,, \end{equation}
up to exponential corrections in $t$ that contribute at small times and
neglecting higher powers of $xt$. Since only $|x|^2$ is relevant, above
we defined $x$ as real and positive.
When the difference $\Delta E = E_{\pi\pi} - E_{\rho}$ is non-zero, we can still perform the sum over $t'$ in Eq.~\eqref{eq:c12}. In this case
the time dependence of the ground state contribution
is $a[\sinh(\Delta E t/2)/\sinh(\Delta E a/2)] e^{-\bar{E}t}$ 
(see, e.g.,
Ref.~\cite{Alexandrou:2013ata}), where the
average energy is defined as
$\bar{E} = \frac{1}{2} (E_{\pi\pi} + E_{\rho})$.
The ground state contribution of the ratio of correlators
can again be used to extract $x$:
\begin{equation} 
\label{eq:rratio}
R(t)\equiv\frac{|C_{12}(t)|}{\sqrt{C_{11}(t)C_{22}(t)}} 
\frac{t\sinh(\Delta E \,a/2)}{a\sinh(\Delta E\, t/2)}
\approx \textmd{const.}+
 xt\,, \end{equation}
where we estimate $\Delta E$ from the exponential decay
of the ratio $C_{11}(t)/C_{22}(t)$ at large (but not too large) times.

We now proceed to estimate
$g_{\rho\pi\pi}$ to assess the reliability of the MMP method.
In Fig.~\ref{fig:directg} we show the resulting ratios
$R(t)$, together with linear fits to the first seven
data points, $6a\leq t\leq 12a$. 
The colour coding of the symbols corresponds to that of
Fig.~\ref{fig:spectra}. The extracted slopes
vary between $51\,\textmd{MeV}$ and $124\,\textmd{MeV}$ with
the smaller slopes corresponding to the larger volume (full symbols),
as one would expect from the naive scaling
with $L^{-3/2}$ of the amplitude $x$ defined in Eq.~\eqref{eq:c12}.
This scaling is also consistent with Eq.~\eqref{eq:g2},
where the combination $x^2L^3$ appears. For the largest slope
$x\approx 124\,\textmd{MeV}$ and $t=12a\approx 0.86\,\textmd{fm}$,
we obtain $xt\approx 0.54$. Indeed, around this Euclidean time higher
order corrections in $xt$ become relevant, while for the large
volume data sets, where the slopes are smaller, the linear behaviour persists
for much longer. We see no indication of exponential
corrections towards small times.

In Table~\ref{tab:gestimates}
we show the results for $x$ and the derived couplings,
where the errors are purely statistical.
More details on the momenta and interpolators used
can be found in Table~\ref{tab:irreps}. The entries of
Table~\ref{tab:gestimates}
are ordered in terms of decreasing $\Delta E$, where
we find that a smaller $\Delta E$ corresponds to
a smaller $E_{cm}$ (and a smaller phase shift $\delta$),
see Fig.~\ref{fig:phaseshift}. Naively, the $T_1$ and $E$ irreps
on the $N_s=48$ lattice should give the most reliable results
as these are closest to the resonance and best matched in terms
of a small $\Delta E$.
However, only the values from the $A_1$ irreps are in agreement
with the result from our L\"uscher-type scattering analysis. We remark that in
terms of the kinematics the $B_1$ irrep is similar to $A_1$, except
for the orientation of the $\rho$ spin relative to the lattice momentum
$\mathbf{K}=(0,1,1)$. These pairs of irreps are also close to each other
in terms of their $\Delta E$ values. Nevertheless, the results from the
$B_1$ irrep differ substantially from the expectation.

Using the L\"uscher method~\cite{Luscher:1990ux}
has the advantage that we can
directly determine the phase shift, without relying on a BW
parametrization or introducing an effective coupling $g_{\rho\pi\pi}$.
Moreover, the systematics can be controlled, while the MMP
method~\cite{McNeile:2002az,McNeile:2002fh}
relies on several approximations that cannot be tested easily.
However, the statistical errors are smaller using the MMP
method than in our full fledged scattering analysis.
In principle we did not even have to evaluate the box diagram
in the upper row of Fig.~\ref{fig:contract} as formally
this is of order $x^2$, beyond the first order perturbative ansatz.
While it is encouraging that the couplings obtained are
of sizes similar to the correct result, they scatter substantially
between volumes and representations.
Therefore, we have to assume a systematic uncertainty of the
MMP method for $\rho$ decay on our volumes of about 50\%,
in terms of the coupling $g_{\rho\pi\pi}$.

\subsection{Comparison to previous results}
\begin{figure}
\includegraphics[width=0.48\textwidth]{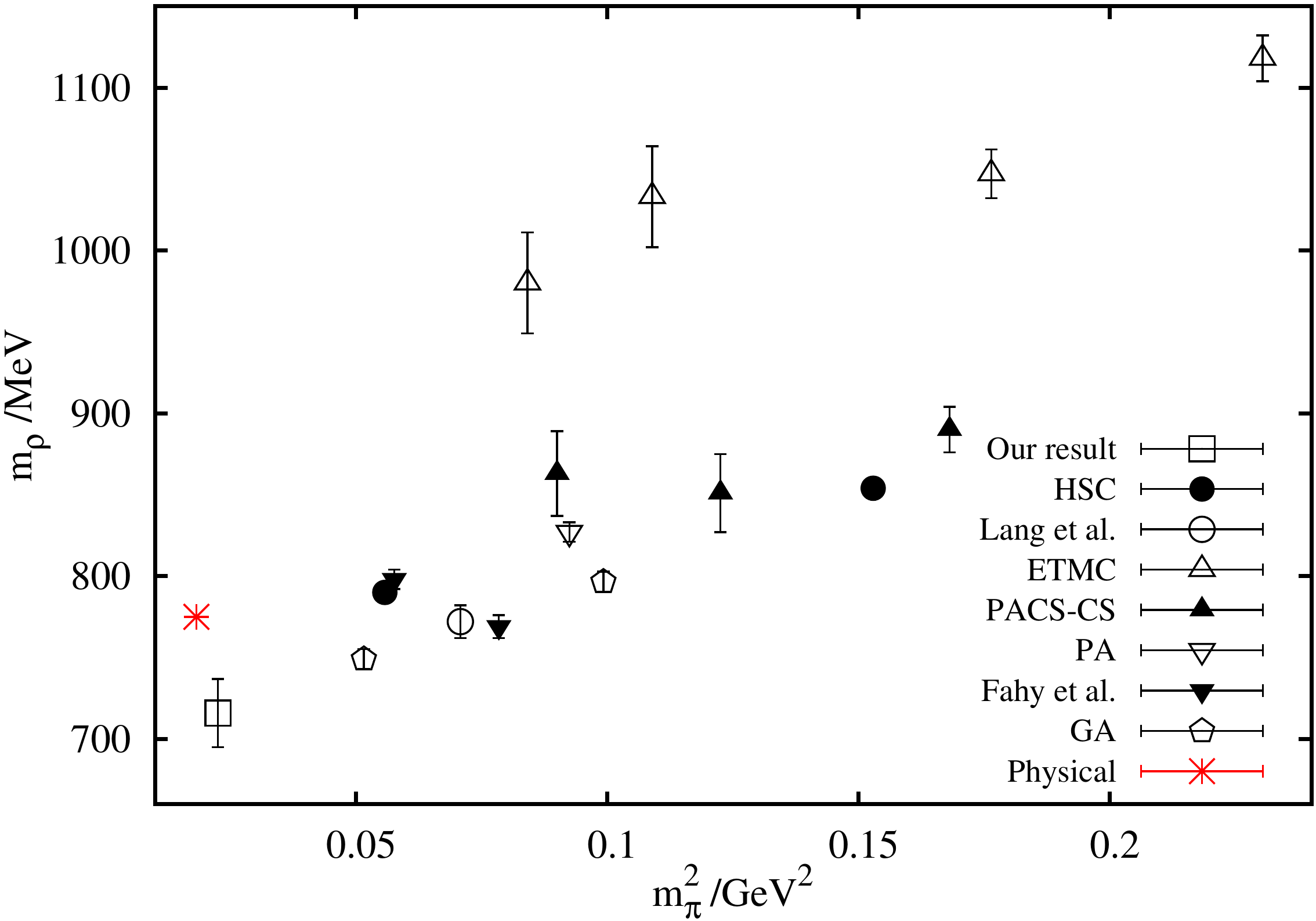}
\caption{$\rho$ resonance masses from this (leftmost open square) and previous lattice calculations by the Hadron Spectrum Collaboration (HSC)~\cite{Dudek:2012xn,Wilson:2015dqa}, Lang et al.~\cite{Lang:2011mn}, ETMC~\cite{Feng:2010es}, PACS-CS~\cite{Aoki:2007rd,Aoki:2011yj}, Pelissier and Alexandru (PA)~\cite{Pelissier:2012pi}, Bulava et al.~\cite{Bulava:2015qjz} and Guo and Alexandru (GA)~\cite{Guo:2015dde}. The physical value is also plotted~\cite{pdg}.
Open symbols correspond to simulations
with $N_f=2$ sea quark flavours, full symbols to $N_f=2+1$.
In none of the cases was the continuum limit taken and no
study includes systematic errors.}
\label{fig:compmrho}
\end{figure}

In Fig.~\ref{fig:compmrho}, we compare our results
on the $\rho$ meson mass,
extracted from the phase shift position $\delta = \pi/2$ of the BW fit
to various results from the literature~\cite{Aoki:2007rd,Feng:2010es,Lang:2011mn,Aoki:2011yj,Pelissier:2012pi,Dudek:2012xn,Wilson:2015dqa,Bulava:2015qjz,Guo:2015dde}.
These results were obtained using different methods, lattice actions,
lattice spacings and $N_f=2$ (open symbols) as well as $N_f=2+1$ (full symbols)
sea quark flavours. In none of the cases was
a continuum limit extrapolation attempted and we only show our
statistical error as the errors of the other data do not contain systematics.
In most of these cases BW masses are quoted, which is why we compare
these to our BW mass.
In Refs.~\cite{Hanhart:2008mx,Pelaez:2010fj} next-to-leading
order (NLO) and next-to-next-to-leading order (NNLO) chiral perturbation theory,
combined with the inverse amplitude method,
are used to predict the pion mass dependence of $m_\rho$.
The quality of the available lattice data does not yet allow for a detailed
comparison. The general trend seen in the majority of lattice calculations
qualitatively agrees with a linear dependence of $m_{\rho}$ on
$m_{\pi}^2$, as suggested by leading order chiral perturbation theory,
however, there are notable outliers.

\begin{figure}
\includegraphics[width=0.48\textwidth]{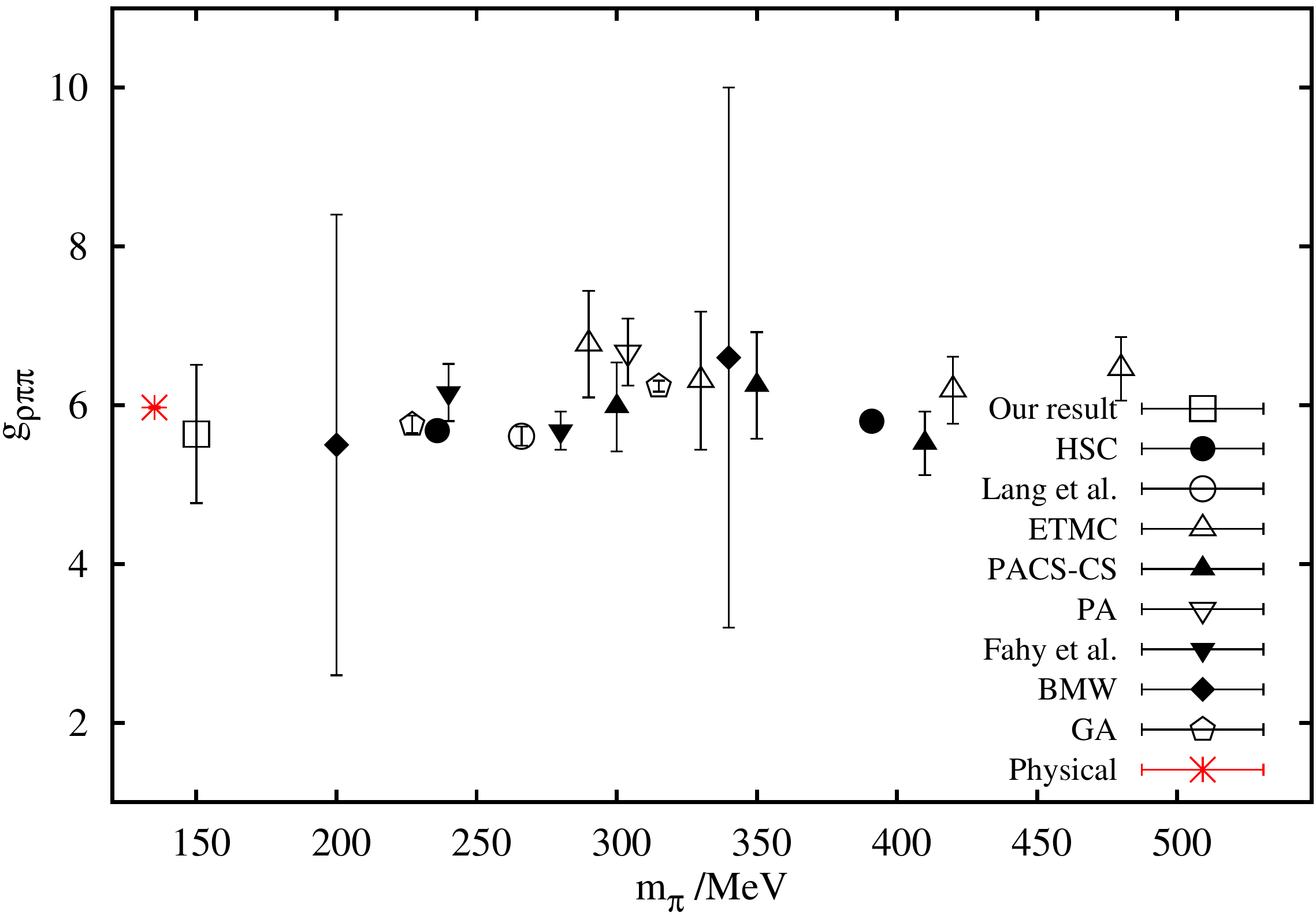}
\caption{Breit-Wigner couplings from various lattice calculations (Hadron Spectrum Collaboration (HSC)~\cite{Dudek:2012xn,Wilson:2015dqa}, Lang et al.~\cite{Lang:2011mn}, ETMC~\cite{Feng:2010es}, PACS-CS~\cite{Aoki:2007rd,Aoki:2011yj}, Pelissier and Alexandru (PA)~\cite{Pelissier:2012pi}, Bulava et al.~\cite{Bulava:2015qjz}, BMW-c~\cite{Frison:2010ws} and Guo and Alexandru (GA)~\cite{Guo:2015dde}) and that extracted from the experimentally measured $\rho$ meson width~\cite{pdg}.
Open symbols correspond to $N_f=2$ results, full symbols to $N_f=2+1$.}
\label{fig:compg}
\end{figure}

In Fig.~\ref{fig:compg} we
show the coupling $g_{\rho\pi\pi}$, obtained
in Refs.~\cite{Aoki:2007rd,Feng:2010es,Frison:2010ws,Lang:2011mn,Aoki:2011yj,Pelissier:2012pi,Dudek:2012xn,Wilson:2015dqa,Bulava:2015qjz,Guo:2015dde}.
Up to $m_{\pi}\approx 400\,\textmd{MeV}$,
Ref.~\cite{Pelaez:2010fj} expects the
coupling $g_{\rho\pi\pi}$ to decrease (increase) by about 5\% at NLO (NNLO),
as a function of the
pion mass, i.e., within the accuracy of their approach, $g_{\rho\pi\pi}$
is constant and the reduction of the decay width is purely due to
phase space. An almost constant behaviour is also suggested by the
Kawarabayashi-Suzuki-Riazuddin-Fayyazuddin
relation~\cite{Kawarabayashi:1966kd,Riazuddin:1966sw},
$g_{\rho\pi\pi}\approx m_{\rho} /f_{\pi} \approx 5.96$,
where $f_{\pi}=\sqrt{2}F_{\pi}
\approx 130\,\textmd{MeV}$ at the physical point.
In Fig.~\ref{fig:compg}, indeed, the lattice values
for pion masses up to $m_\pi \approx 470\,\textmd{MeV}$ are all
around this coupling (which is indistinguishable
from the physical coupling $g_{\rho\pi\pi}\approx 5.93$, also shown
in the figure). However, the noise increases significantly, closer
to the physical pion
mass, so $g_{\rho\pi\pi}$ can be extracted much more accurately
at large quark masses.
Again note that the lattice results were obtained at
different lattice spacings with different actions and have quite different
systematics.

For $K\pi$ scattering only a few previous lattice studies exist.
At $m_{\pi}\approx 150\,\textmd{MeV}$ and at our lattice spacing,
we find (Eqs.~\eqref{eq:massk} and \eqref{eq:coupling})
$m_{K^*}=868(13)(26)\,\textmd{MeV}$ and $g_{K^*\pi\pi}=4.79(49)$.
Note that in experiment $m_{K^*}\approx 896\,\textmd{MeV}$ and
$g_{K^*K\pi}\approx 5.39$.
The Hadron Spectrum Collaboration~\cite{Wilson:2014cna}
reports $m_{K^*} = 933(1)\,\textmd{MeV}$
and $g_{K^*K\pi} = 5.72(52)$ at a pion mass of $391\,\textmd{MeV}$.
Prelovsek et al.~\cite{Prelovsek:2013ela} use $m_\pi = 266\,\textmd{MeV}$ and obtain $m_{K^*} = 891(14)\,\textmd{MeV}$ and $g_{K^*K\pi} = 5.7(1.6)$
while Fu and Fu~\cite{Fu:2012tj} find $m_{K^*} = 1014(27)\,\textmd{MeV}$ and
$g_{K^*\pi\pi} = 6.38(78)$, using a lattice spacing of $0.15\,\textmd{fm}$
and a pion mass of $240\,\textmd{MeV}$.

\section{Conclusions}
\label{sec:conclude}
In summary, we have demonstrated the feasibility of computing
resonance scattering parameters at a nearly physical pion mass.
In particular, we computed the $p$-wave scattering phase shifts
for $\pi\pi$ scattering in the $I=1$ channel and $K\pi$ in the $I=1/2$ channel.
From these, we extracted the
masses and couplings
$m_\rho = 716(21)(21)\,\textmd{MeV}$,
$\Gamma_\rho = 113(35)(3)\,\textmd{MeV}$, $m_{K^*} = 868(13)(26)\,\textmd{MeV}$
and $\Gamma_{K^*} = 30(6)(1)\,\textmd{MeV}$.
The masses are lower than the experimental ones,
$m_{\rho}\approx 775\,\textmd{MeV}$,
$m_{K^*}\approx 896\,\textmd{MeV}$, and at least the
width of the  $K^*$ meson is underestimated too,
in part due to a 10\% heavier than physical pion.
The values from experiment are:
$\Gamma_{\rho}\approx 148\,\textmd{MeV}$,
$\Gamma_{K^*}\approx 47\,\textmd{MeV}$~\cite{pdg}.
The second errors reflect an overall
scale uncertainty of 3\%~\cite{Bali:2012qs}.
While for the $\rho$ meson mass and width this error can
be added in quadrature to the statistical one,
for the $K^*$ parameters it is not straightforward to
account for this uncertainty as our strange quark mass was tuned, assuming
$a^{-1}=2.76\,\textmd{GeV}$. It is clear that we undershoot
the experimental $\rho$ resonance mass by about two standard deviations,
which indicates that not all systematics have been accounted for,
in particular only one (albeit small) lattice spacing was realized.
The corresponding positions of the resonance poles in the second
Riemann sheet from analytical
continuation are shown in Tables~\ref{tab:rho} and \ref{tab:kstar} and,
at our present level of error, these cannot be distinguished from
the above Breit-Wigner fit results.

The stochastic one-end source method we have used is cheaper compared to
other methods~\cite{Peardon:2009gh,Morningstar:2011ka,Fahy:2014jxa}, as
long as the set of kinematic points (and interpolators) is suitably restricted. 
In our calculation, we were able to recycle many propagators, by keeping
one of the momenta, $\mathbf{p}_1$, fixed.
The number of inversions required is given in Eq.~\eqref{eq:cost}
and the cost of including additional momenta is large.
This is a limitation in particular for larger volumes, when the density
of states increases and the use of multiple two-particle interpolators
cannot be avoided. We remark, however, that our larger volume
with a linear lattice extent $64a\approx 4.6\,\textmd{fm}$ is not at all
small considering present-day standards in lattice scattering computations.

An alternative approach is the distillation method~\cite{Peardon:2009gh},
which has been used in several other scattering calculations \cite{Dudek:2012xn,Lang:2011mn,Wilson:2015dqa}.
This method does not suffer from a large computational overhead when
including additional momenta as time-slice-to-all propagators
(perambulators) are used in constructing the correlators.
However, this method is not very well suited to large volumes as
the number of vectors required increases in proportion to $L^3N_t$
and the cost of contractions also scales with a power of the number of vectors.
Combining this method with stochastic estimates~\cite{Morningstar:2011ka}
may ultimately not change this scaling behaviour but may make realistic
lattice sizes accessible. Indeed, this stochastic distillation method has been
successfully employed for $\pi\pi$
scattering~\cite{Fahy:2014jxa,Bulava:2015qjz}, where the number of solves
used in Ref.~\cite{Bulava:2015qjz}
is not much higher than ours. It will be very interesting to see
if such calculations can be pushed towards small quark masses, large volumes 
and time distances of about $1\,\textmd{fm}$ that allow for a reliable
extraction of energy levels. Stochastic distillation was also successfully used
to study $DK$ scattering~\cite{Mohler:2013rwa,Lang:2014yfa,Lang:2015hza}.

Our calculation is performed at a single lattice spacing and it is
not possible to quantify the size of discretization effects.
For the action we use, these are of $\mathcal{O}(a^2)$ and it is
unlikely at our lattice spacing $a \approx 0.071\,\textmd{fm}$
that they are much larger than our 3\% scale uncertainty.
Limited information for the $\mathcal{O}(a^2)$ accurate twisted
mass action can be extracted from the results for
the $\rho$ meson mass given in Ref.~\cite{Burger:2013jya}.
In this study of the hadronic vacuum polarization contribution
to $(g-2)_\mu$, the correlators for vector mesons are calculated
only using a one-particle interpolator for several ensembles with different
lattice spacings and (larger than physical) pion masses.
The mass of the $\rho$ is then found by treating it as a stable particle
and the results obtained show no significant dependence on the lattice spacing.
We therefore assume that the 3\% scale uncertainty and the
10\% larger than physical pion mass are dominant systematics but we cannot
exclude other sources of error, in particular lattice spacing effects or the
omission of the strange quark from the sea.

In Figs.~\ref{fig:compmrho} and \ref{fig:compg} we compare our results
on the $\rho$ meson mass and coupling to
those of other lattice studies that were carried out at larger pion masses.
The coupling $g_{\rho\pi\pi}$ appears to be remarkably independent of the quark
mass and also robust against other systematics.

Future work will extend the present study to $N_f = 2+1$ flavour configurations,
including several lattice spacings, to enable a continuum limit extrapolation.
Working close to the physical pion mass is particularly
valuable for simulations of scattering processes involving states
that are near to thresholds, e.g., $X(3872)$ and $D\overline{D}^*$
or $D_{s0}(2317)$ and $DK$, where the gap relative to the threshold
strongly depends on the light quark mass.

\acknowledgments
This work was supported by the Deutsche Forschungsgemeinschaft grant SFB/TRR 55.
The authors gratefully acknowledge the Gauss Centre for Supercomputing e.V. (\url{http://www.gauss-centre.eu}) for granting computer time on
SuperMUC at Leibniz Supercomputing Centre (LRZ, \url{http://www.lrz.de})
for this project.
The {\sc Chroma} \cite{Edwards:2004sx} software package was used, along with
the locally deflated domain decomposition solver implementation of {\sc openQCD} \cite{ddopenqcd}.
The ensembles were generated primarily on the SFB/TRR~55 QPACE
computer~\cite{Baier:2009yq,Nakamura:2011cd}, using
{\sc BQCD}~\cite{Nakamura:2010qh}.
G.~S.~Bali and S.~Collins acknowledge the hospitality of the
Mainz Institute for Theoretical Physics (MITP) where a significant
portion of this article was completed. We thank Simone Gutzwiller~\cite{simonethesis} and Tommy Burch for preparatory work,
Andrei Alexandru for discussion
and Benjamin Gl\"a\ss{}le for software
support.
\bibliography{things}
\end{document}